# The Mass Distribution of the Lens Galaxy in MG1131+0456

G. H. Chen
Department of Physics and Research Laboratory of Electronics, Massachusetts Institute of Technology, Cambridge, MA 02139

C. S. Kochanek
Harvard-Smithsonian Center for Astrophysics and Department of Astronomy, Harvard College Observatory, Cambridge MA 02138

J. N. Hewitt
Department of Physics and Research Laboratory of Electronics, Massachusetts Institute of Technology, Cambridge, MA 02139

## ABSTRACT

We present the results of modeling 5, 8 and 15 GHz maps of the gravitational lens MG1131+0456 using the LensClean algorithm. Two models for the mass distribution in the lens were fit: a de Vaucouleurs model and a model with the two-dimensional potential $\phi = b^{2-\alpha}(r^2 + s^2)^{\alpha/2}/2\alpha$, both in an external shear field. The best fit de Vaucouleurs model has an effective radius of $R_e = 0\farcs83 \pm 0\farcs13$, larger than that measured at optical wavelengths. The best fit "$\alpha$ model" has $\alpha = 0.6 \pm 0.2$ and a core radius of $s = 0\farcs19 \pm 0\farcs07$. An isothermal model ($\alpha = 1$) is inconsistent with the data. The best $\alpha$ model fits the 8 GHz map significantly better than the best de Vaucouleurs model, and it is a good fit at all three frequencies. Although the peak residuals in the best reconstructions are less than 5% of the map peak and it is difficult to distinguish visually the original image and the reconstructions, none of the models is statistically consistent with the data. Experiments on the effects of using Clean maps suggest that systematic errors are not the source of the discrepancy. We believe the primary problem is that an external shear is an inadequate model for the angular structure of the lens galaxy. The finite core radius in the models is required by the structure of the extended ring, so we confirm that the central component in the 8 GHz map is a central, lensed image of the source rather than the lens galaxy. Two galaxies $3''$ from the ring add an external shear to the lens model that is within $3°$ of the fitted model. The alignment is remarkable, but the two galaxies cannot account for the magnitude of the shear unless the effective ratio of their luminosity to the primary lens galaxy is twenty-five times larger than the measured ratio. The model lens positions, the position angle of the shear, and the mass interior to the ring are determined very precisely by the lens models. The time delay between the compact components is predicted with an uncertainty we estimate to be at most $\pm 9\%$; the formal uncertainty is $\pm 4\%$.

## 1. Introduction

The mass distribution in early type galaxies is still an open and important problem in astronomy because of the limitations of dynamical techniques. Dynamical observations are consistent with a constant mass-to-light ratio for the inner regions of ellipticals (eg. de Zeeuw & Franx 1991, van der Marel 1991), but they are also consistent with models containing dark matter (eg. Bertin et al. 1992, Saglia et al. 1993). Some, possibly atypical, ellipticals have dynamical tracers like X-ray halos (eg. Fabbiano 1989) that imply much larger masses for the outer regions. In addition to the uncertainties in the radial structure of the galaxies there is evidence in NGC720 that the asymmetries in the mass and the light are misaligned (Buote & Canizares 1994). Gravitational lensing depends only on the gravitational potential of the lens galaxy, so lenses should provide new and more direct evidence on the mass distribution of galaxies.

The possibility of using gravitational lensing to investigate the structure of galaxies was realized (Bourassa & Kantowski 1975) even before the discovery of the first gravitational lens system (Walsh et al. 1979). However, the high expectations for using gravitational lens systems to study the mass distribution of the lens galaxies have yet to be fulfilled. While the lenses can measure some parameters of the lens very accurately, such as the mass inside the ring of images or the average ellipticity at the ring, there is still no convincing determination of the radial distribution of matter in any lens galaxy. Statistical studies of lens surveys (Kochanek 1993, Maoz & Rix 1993) suggest that constant mass-to-light ratio models for E/S0 galaxies are incompatible with the statistics of gravitational lenses, while isothermal distributions are consistent. Although statistical studies are valuable in obtaining collective information on lens galaxies, the conclusions are sensitive to selection effects and other shortcomings of the statistical model. Most lens models that fit the systems consisting of two or four unresolved images consist of simple parameterized models for the mass distribution such as isothermal spheres, de Vaucouleurs models, King models, Plummer models, and modified Hubble models. The problem with these systems is that the paucity of constraints and the restriction of the images to a limited range of radii from the center of the lens means that lens models of these systems generally provide little or no constraint on the radial distribution of matter and cannot distinguish between different radial mass distributions (Kochanek 1991, Wambsganss & Paczyński 1994).

In the case where the source of the lens is extended, as in the radio rings (see Patnaik 1994 for a review) and the cluster arcs (see Soucail & Mellier 1994 for a review), the number of constraints is greatly increased. In particular, the radio rings frequently have emission over an extended range of radii, so we should be able to determine the radial distribution of matter by modeling the lens. The difficulty in systematically and accurately modeling an extended lens has slowed progress, and only MG1131+0456 (Kochanek et al. 1989), PKS1830-211 (Kochanek & Narayan 1992), Cl0024+1654 (Wallington & Kochanek 1994), and MG1654+134 (Kochanek 1994) have been studied using true inversion methods. The existing model of MG1131+0456 used an inversion method that did not account for the instrumental resolution of the VLA, which severely limits the accuracy of the models. In this paper we reexamine MG1131+0456 using new radio data (Chen & Hewitt 1993) and a more sophisticated inversion algorithm.

MG1131 consists of five radio components: a radio ring, two compact components (A and B) abutting the ring, one faint extended component outside the ring (C), and one unresolved component (D) at the center of the ring (see fig. 1). The system has been imaged at radio frequencies of 5, 8, 15, and 22 GHz (Chen & Hewitt 1993), so there is a wealth of structural information available. The source is evident at both optical and infrared wavelengths (Hammer et al. 1991, Annis 1992, Larkin et al. 1994). At infrared wavelengths there is an excess of galaxies within $20''$ of MG1131, suggesting there may be a cluster of galaxies in the field. Statistical models for the lens redshifts by Kochanek (1992) give median lens redshifts of $z_l = 0.3$, 0.5, and 0.6 for source redshifts of $z_s = 1$, 2, and 3.

The system was modeled by Kochanek et al. (1989) using the "Ring Cycle" algorithm to invert the 15 GHz maps from Hewitt et al. (1988) under the assumption that the flux densities in the image represent the true surface brightness. They fitted an elliptical isothermal potential model for the galaxy to obtain a convincing model for the formation of the ring, although the modeled ring always appeared to be slightly less elliptical than the observed ring. The assumption in the algorithm of a surface brightness map is a severe shortcoming, because the resolution of even the 15 GHz maps has a perceptible effect on the structure of the image. Lower frequency maps (5 GHz) simply could not be modeled because of the breakdown of these assumptions. Kochanek & Narayan developed the "LensClean" algorithm to address these difficulties by using a generalization of the Clean algorithm (Högbom 1974, Clark 1980) to invert lensed sources taking into account the instrumental resolution.

In this paper we model the 5, 8, and 15 GHz radio images of MG1131 summarized in Table 1 using the LensClean algorithm and a family of lens models. We chose not to study the 22 GHz data in any detail because only the compact components (A and B) are detected at this frequency with high signal-to-noise. Because the spectral indices of the compact components and the extended components are different, maps at different frequencies emphasize different features of the system. In §2 we describe the model potentials for the lens, the optimization procedure, the statistical error analysis for the models, and the systematic errors that arise from using a Clean map as the starting point for the inversions.

In §3 we present the models for the 8 GHz images, address the role of the central component D, and discuss the effects of two nearby galaxies on the lensing potential. In §4 we show that the best fit models for the 8 GHz images are also consistent with the 5 GHz and 15 GHz images. In §5 we discuss the implications of the models.

## 2. Method

### 2.1. Parameterized Potentials

We use two simple, non-singular circular potentials in an external shear field to describe the two dimensional Newtonian potential of the lens galaxy in polar coordinates $(r,\ \theta)$ centered at the position of the lens. The angular coordinate $\theta$ is measured from west to north, and the coordinates $(x, y)$ represent a right-handed coordinate system with $x = r\cos\theta$ and $y = r\sin\theta$. We first consider the de Vaucouleurs (1948) model, since it is the prototypical constant mass-to-light ratio model of an elliptical galaxy. The monopole structure of the de Vaucouleurs (1948) model depends only on the total mass of the galaxy and the effective radius of the surface density $R_e$. Next we consider a family of potentials, which we call the $\alpha$ model, that is characterized by a central core radius $s$ and a power-low slope $\alpha$. This class of models allows us to explore a wide range of shapes for the monopole of the lens.

If the two-dimensional lensing potential of the lens is $\phi$, then the thin-screen lens equation for the deflection of rays is

$$\mathbf{u} = \mathbf{x} - \bigtriangledown\phi(\mathbf{x}), \tag{2-1}$$

where $\mathbf{u}$ and $\mathbf{x}$ are angular coordinates on the source and image plane (eg. Schneider, Ehlers & Falco 1992). The inverse magnification tensor on the image plane is the Hessian matrix of the transformation

$$M_{ij}{}^{-1} = (\delta_{ij} - \partial_i\partial_j)\phi, \tag{2-2}$$

and the total magnification is the inverse of the determinant of the Hessian matrix $M^{-1} = |M_{ij}^{-1}|$. The surface density of the lens $\Sigma$ is determined by $\nabla^2\phi = 2\Sigma/\Sigma_c$ where $\Sigma_c = c^2 D_{OS} D_{OL}/4\pi G D_{LS}$ is the critical surface density for lensing (in units of grams per square angle), and the $D_{ij}$ are angular diameter distances between the observer, lens, and source redshifts. Neither the source nor the lens redshift are known for MG1131, so we cannot convert the measured angular parameters of the lens model into physical masses and lengths with any certainty.

The mass interior to radius $r$ of the de Vaucouleurs model is $M(<r) = MF(r/R_e)/2$ where $M$ is the total mass of the galaxy, the function $F$ is

$$F(r/R_e) = \left[\int_0^{r/R_e} uI(u)du\right]\left[\int_0^1 uI(u)du\right]^{-1} \qquad 0 \leq F \leq 2 \tag{2-3}$$

(Maoz & Rix 1993) and $I(x) \propto \exp(-7.67(x^{1/4}-1))$ is the surface brightness at radius $x = r/R_e$ from the center of the galaxy. The lens equation produced by this monopole in an external shear of amplitude $\gamma$ and position angle $\theta_\gamma$ is

$$\mathbf{u} = \mathbf{x} - bR_eF\left[\frac{r}{R_e}\right]\frac{\mathbf{x}}{r^2} - \gamma\begin{pmatrix}\cos 2\theta_\gamma & \sin 2\theta_\gamma \\ \sin 2\theta_\gamma & -\cos 2\theta_\gamma\end{pmatrix}\begin{pmatrix}x \\ y\end{pmatrix} \tag{2-4}$$

where the characteristic deflection scale is $b = [D_{LS}/(D_{OL}D_{OS})][(2GM/c^2R_e)]$. The total mass of the galaxy is

$$M = \frac{D_{OS}D_{OL}}{D_{LS}}\frac{c^2R_eb}{2G} = 1.5h^{-1}R_eb\left[\frac{D_{OS}D_{OL}}{2r_HD_{LS}}\right] \times 10^{12}M_\odot \tag{2-5}$$

where $R_e$ and $b$ are in arc seconds, the Hubble constant is assumed to be $H_0 = 100h$ km s$^{-1}$ Mpc$^{-1}$, and $r_H = c/H_0$ is the Hubble radius. For the median lens redshifts given above, the combination of the angular diameter distances in brackets equals 0.16, 0.22, and 0.23 for source redshifts of $z_s = 1$, 2, and 3.

The two-dimensional lensing potential for the $\alpha$ model is

$$\phi = \frac{b^{2-\alpha}(r^2+s^2)^{\alpha/2}}{\alpha} + \frac{1}{2}\gamma r^2\cos 2(\theta-\theta_\gamma), \tag{2-6}$$

where $b$ is the tangential critical radius of the lens if $s = 0$, and $s$ is the core radius (Blandford & Kochanek 1987). The shape exponent $\alpha$ determines how rapidly the density declines with radius. If $\alpha = 1$ the model approaches an isothermal sphere (for $r \gg s$, $\phi \propto r$, surface density $\Sigma \propto 1/r$), while if $\alpha = 0$, the potential reduces to a Plummer model (for $r \gg s$, $\phi \propto \ln r$, $\Sigma \propto 1/r^4$). The Kochanek et al. (1989) models correspond to $\alpha = 1$ and a different angular structure. The surface mass density of the $\alpha$-model is

$$\Sigma = \frac{1}{2}\Sigma_c\nabla^2\phi = \frac{1}{2}\Sigma_cb^{2-\alpha}(r^2+s^2)^{(\alpha-2)/2}\left[\frac{2s^2+\alpha r^2}{r^2+s^2}\right], \tag{2-7}$$

and the mass inside projected radius $r$ is

$$M(<r) = \pi\Sigma_cb^{2-\alpha}r^2(r^2+s^2)^{(\alpha-2)/2} = 7.4h^{-1}b^{2-\alpha}r^2(r^2+s^2)^{(\alpha-2)/2}\left[\frac{D_{OS}D_{OL}}{2r_HD_{LS}}\right] \times 10^{11}M_\odot \tag{2-8}$$

for $b$ and $s$ in arc seconds.

The $\alpha$ model is analytically tractable and we can derive several useful scalings of the lensing properties of the model that explain the parameters determined by the full inversion. The shape and size of the ring are set by the semimajor and semiminor axes of the tangential critical line

$$r_\pm^2 = b^2(1 \mp \gamma)^{2/(\alpha-2)} - s^2, \tag{2-9}$$

which is approximately equal to $r_\pm \simeq b\,[1 \pm \gamma(2-\alpha)^{-1} - \beta^2/2]$ if $\gamma \ll 1$ and $\beta = s/b \ll 1$. The average ring radius is $\langle r \rangle = (1/2)(r_+ + r_-) \simeq b(1 - \beta^2/2)$ and the average ring ellipticity is $\epsilon = 1 - r_-/r_+ \simeq 2\gamma/(2-\alpha)$. If the core radius is small ($\beta \ll 1$) then the average ring radius determines the parameter $b$, and the ellipticity of the ring determines the dimensionless shear $\gamma$ and the exponent $\alpha$. For a circular lens the average density inside the tangential critical line is equal to the critical density $\Sigma_c$, so the mass interior to the ring or the critical radius, $M(<\langle r \rangle) = \pi \langle r \rangle^2 \Sigma_c$, is independent of the lens parameters.

The value of the core radius is constrained by the flux of the central (or "odd") image, the size of the multiply imaged region, and the slope of the deflection angle at the ring radius. The flux of the central image depends on the magnification at the center of the lens ($r = 0$)

$$M_0 = \left[1 - \left(\frac{b}{s}\right)^{2-\alpha}\right]^{-2} \simeq \beta^{2(2-\alpha)} \tag{2-10}$$

for $\beta \ll 1$. Since $b$ is largely fixed by the diameter of the ring, the core radius controls the flux of the central image and a larger core radius $s$ produces a brighter central image. There are no simple analytic models for the size of the multiple image region, but the trends can be understood from the behavior of the bending angle and the peak deflection of the lens. The monopole deflection of the lens is $b^{2-\alpha} r (r^2 + s^2)^{(\alpha-2)/2}$ which peaks at $r^2 = s^2/(1-\alpha)$. When $\alpha < 1$ a smaller core radius increases the peak deflection and expands the multiply imaged region.

The shear in these models is caused by matter outside the region being modeled. The axis $\theta_\gamma$ is the major axis of the shear (inward deflections of rays) measured from West to North and $\gamma$ is its dimensionless strength. If the quadrupole mass distribution of the material generating the shear is $\Sigma_2(r)\cos 2(\theta - \theta_\Sigma)$ with $\Sigma_2(r) > 0$, then the major axis of the matter distribution is perpendicular to the major axis of the shear ($\theta_\gamma = \theta_\Sigma + \pi/2$) and the magnitude of the shear is

$$\gamma = \int_r^\infty \frac{dr}{r} \frac{\Sigma_2}{\Sigma_c} \tag{2-11}$$

where $\Sigma_c$ is the critical density for lensing. For an ellipsoidal surface density $\Sigma[r^2(1+\epsilon\cos 2\theta)]$ the shear is just $\gamma = (\epsilon/2)(\Sigma(r)/\Sigma_c)$.

### 2.2. Optimizing the Lens Parameters in Multi-Dimensional Space

The de Vaucouleurs (1948) model has six free parameters (the effective mass $b$, the effective radius $R_e$, the lens position ($x_l$, $y_l$), the dimensionless shear $\gamma$, and the orientation of the shear $\theta_\gamma$) and the $\alpha$ model has seven free parameters (the critical radius $b$, the lens position ($x_l$, $y_l$), the dimensionless shear $\gamma$, the orientation of the shear $\theta_\gamma$, the exponent $\alpha$, and the core radius $s$). For a given set of lens parameters, LensClean finds the best fit source model and the residual flux in the image after subtracting the model from the data. By minimizing these residuals as a function of the free parameters we find the best fit lens model.

We want to extract not only the best fit model for all free parameters, but also the dependence of the errors on the primary structural parameters of the models: $R_e$ for the de Vaucouleurs model, and $\alpha$ and $s$ for the $\alpha$ model. These parameters determine the shape of the monopole and have the greatest uncertainty, while the remaining five parameters are very strongly constrained by the geometry of the ring. We use the following procedures to isolate these parameters and to ensure that we are avoiding local extrema in the error surface:

(1) For a model with fixed structural parameters ($R_e$ or $\alpha$ and $s$) we test a series of models optimizing the variables $b$, $\theta_\gamma$, and $\gamma$ on a grid of fixed lens positions $x_l$ and $y_l$. The lens position is reasonably well determined from the image geometry, so we examine a region approximately $0''.80$ by $0''.64$ centered on the position of component D. The initial values of the variables are set to fit the positions and fluxes of the compact components. The optimization of the remaining parameters is rapid and well determined. The best solutions on the grid of positions are used as the initial data for a final optimization to determine the best model for the current values of the structural parameters.

(2) We do a series of these models with varying length scales $R_e$ or $s$ (at fixed $\alpha$) to determine the best fit scale radii and their allowed ranges.

(3) For the $\alpha$ model we repeat these procedures for the range from $\alpha = 0.3$ to $\alpha = 1$. Outside this range the residuals begin to rise steeply.

(4) Once we have isolated reasonable models for the 8 GHz image of the lens we repeat the procedures for the 5 GHz and 15 GHz images.

All the intermediate solutions and their residuals are kept as a database for estimating the errors in the parameter estimates. To interpret the residuals we need to examine both random and systematic sources of error in the inversion method.

### 2.3. Goodness of Fit and Error Estimation

LensClean makes a $\chi^2$ fit of the reconstructed image to the input image, so the fundamental measure of the error is the mean square difference between the two images. To construct the $\chi^2$ statistic, we must define the number of degrees of freedom, $N_{dof}$, the portions of the map that contribute to the error, and the noise level. The noise level is at least the measured noise in the map $\sigma_0$ (rms noise per pixel), although systematic errors (see §2.4) may lead to a higher effective noise level. Let $N_{pix}$ be the number of pixels of size $\Delta x$ in the map, and $N_{mult}$ be the number of multiply imaged pixels in the map. Let the rms residual per pixel over the entire map be $\sigma_{tot}$ and the rms residual per pixel over the multiply imaged region be $\sigma_{mult}$. The definition of $\chi^2$ and $N_{dof}$ depends on the effects of finite resolution and the subtraction of the residuals from the original Clean.

The original Clean of the map produces a components map and a residual map. The components are convolved with a Gaussian Clean beam and the residuals are added to produce the final Clean image. If the Clean is deep enough, the residual map is uncorrelated with the dirty beam and represents random or systematic noise in the measured visibilities. When we LensClean the map using the Clean beam, we can subtract most of the original residuals because of the differences between the compact Clean beam and the complicated dirty beam. LensCleaning the Clean map produces negligible residuals in the singly imaged region of the lens. In the multiply imaged region, the noise at the location of the different images is uncorrelated, so we cannot subtract all the residuals.

The number of independent data points is not the number of pixels, but the number of independent beam elements in the map. This is most easily seen by taking the "data" to be the gridded Fourier components of the map where

$$\chi^2 = \sum_{i=1}^{N_{data}} \frac{(\tilde{I}_{ij} - \tilde{I}^m_{ij})^2}{\sigma_f^2} \tag{2-12}$$

and $N_{data}$ is the number of Fourier components, $\sigma_f$ is the noise associated with each Fourier component, and $\tilde{I}_{ij}$ and $\tilde{I}^m_{ij}$ are the measured and modeled Fourier components. The number of cells occupied by the clean beam in Fourier space is

$$N_{data} = \frac{(\Delta x)^2}{2\pi\sigma_b^2} N_{pix} \tag{2-13}$$

where $\sigma_b$ is the dispersion of the Gaussian beam. The number of independent data points is the same as the area of the map $N_{pix}\Delta x^2$ divided by the effective area of the Gaussian beam $2\pi\sigma_b^2$.

We use Parseval's theorem to relate residuals and errors in the Fourier and map planes,

$$\sum(\tilde{I}_{ij} - \tilde{I}^m_{ij})^2 = N_{pix}\sigma^2_{tot} \qquad \text{and} \qquad N_{data}\sigma^2_f = N_{pix}\sigma^2_o \tag{2-14}$$

where $\sigma^2_{tot}$ is the pixel-to-pixel residual after LensClean. Thus the $\chi^2$ statistic for the total map is

$$\chi^2_{tot} = N_{data}\frac{\sigma^2_{tot}}{\sigma^2_o} = N_{pix}\frac{(\Delta x)^2}{2\pi\sigma^2_b}\frac{\sigma^2_{tot}}{\sigma^2_o}. \tag{2-15}$$

and the $\chi^2$ statistic for just the multiply imaged regions of the map is

$$\chi^2_{mult} = N_{mult}\frac{(\Delta x)^2}{2\pi\sigma^2_b}\frac{\sigma^2_{mult}}{\sigma^2_o}. \tag{2-16}$$

To evaluate the significance of the $\chi^2$, we must also determine the number of degrees of freedom. There are $M$ parameters in the lens model, but for the moment these are not important in estimating $N_{dof}$ because $M$ is small for both models. The LensClean model fits a large number of source components, and the number of source components is the main contribution to $N_{dof}$. Suppose the image we use has "infinite" resolution in the sense that no images of one source point are correlated with the images of another source point by the effects of the beam. A singly imaged source has one measured flux in the image, a triply imaged source has three measured fluxes, and a quintupally imaged source has five measured fluxes. In each case we fit one source flux parameter leaving no degrees of freedom for the singly imaged source, two for the triply imaged source, and four for the quintupally imaged source. If we fit the entire image plane in which fraction $f_1$ is singly imaged, $f_3$ is triply imaged, and $f_5$ is quintupally imaged ($f_1 + f_3 + f_5 = 1$) then the number of degrees of freedom is $N_{dof} = N_{data}(1 - f_1 - f_3/3 - f_5/5) - M \simeq (2/3)N_{mult}$ (Wallington & Kochanek 1994). The number of degrees of freedom is proportional to the area that is multiply imaged.

With the addition of a beam that correlates nearby images, this estimate exaggerates the number of degrees of freedom in the model. For example, the Plummer model ($\alpha = 0$) with a fixed ring radius produces a larger and larger multiply imaged region as the core radius is reduced. In the limit that $s \to 0$, the multiply imaged region becomes infinite, suggesting that $N_{dof} \to \infty$. With finite resolution this result is clearly incorrect. Most of the large multiple image region consists of sources that have one image at a large radius from the lens center with nearly unit magnification, and two strongly demagnified images in the core of the lens. Consider the case where the potential has a singular core. Let there be a pair of sources, each with one image at a large radius and one image in the core. If we simultaneously fit two such image pairs, the previous approximation gives two degrees of freedom because we fit four image fluxes with two source fluxes. However, the beam makes the two images in the core fully correlated (ie. they are the same point) so we are really fitting three image fluxes with two source fluxes giving only one degree of freedom. For $N$ such image pairs, all $N$ images in the core are correlated, so there are only $N+1$ image fluxes available to model $N$ source fluxes, leaving only one degree of freedom. Thus with the addition of a finite beam size, the number of degrees of freedom is not simply proportional to $N_{mult}$, because many of the multiple image systems act as if they were singly imaged.

The tangential critical line effectively separates the inner images from the outermost images and suggests the correct formulation for the number of degrees of freedom given multiply imaging and finite resolution. Suppose we have a symmetric lens that generates only one or three images, and we assign the source flux to fit the images outside the tangential critical line exactly. All the residuals from this procedure are inside the tangential critical line, and the number of degrees of freedom is the number of resolution elements inside the tangential critical line. This model generalizes to simultaneously fitting all the images, and it is approximately correct when we add the five image region. Thus the correct estimate for the number of degrees of freedom in the models is

$$N_{dof} = \frac{A_{tan}}{2\pi\sigma_b^2} - M \qquad (2\text{-}17)$$

where $A_{tan}$ is the area inside the tangential critical line of the lens. $A_{tan}$ has the nice property of being fixed for all reasonable lens models, because all reasonable lens models must have the same average tangential critical line to be able to fit the ring. Thus $A_{tan} = \pi\langle r\rangle^2$ where $\langle r\rangle$ is the average ring radius, with corrections that are second order in the dimensionless shear, $\gamma$. The approximation fails when the size of the tangential critical line is comparable to the size of the Clean beam. Note that a good inversion should have an rms residual for the whole map smaller than the intrinsic noise by the factor $(A_{tan}/A_{map})^{1/2}$, and an rms residual for the multiple image region smaller than the intrinsic noise by the factor $(A_{tan}/A_{mult})^{1/2}$.

Given a $\chi^2$ and $N_{dof}$, and assuming the errors are Gaussian and uncorrelated, we expect $\chi^2 = N_{dof}$ for a good inversion with the value of $\chi^2$ distributed as a $\chi^2$ distribution with $N_{dof}$ degrees of freedom. Since $N_{dof} \gg 1$ the expected standard deviation of the $\chi^2$ from $N_{dof}$ is approximately $(2N_{dof})^{1/2}$. In practice, we tend to find larger values of $\chi^2$ both because of the real limitations in the models and because of systematic error. One source of systematic error is the interpolation and reconstruction error (see §2.4). A second systematic difficulty is that LensClean needs to use an automatic stopping criterion to decide when to stop cleaning. Like normal uses of Clean on extended sources, it is possible to Clean more or less deeply by varying the total number of components in use. We stop the LensClean when a major cycle of the Clean does not reduce the peak residual, so there is still subtractable flux in the map when the procedure stops. We can compensate for the uncertainties in the true noise level and the stopping criterion by renormalizing the errors so that $\chi_r^2 = N_{dof}\chi^2/\chi_{min}^2$, where $\chi_{min}^2$ is the smallest measured value of $\chi^2$. This will systematically *underestimate* the statistical significance of differences among models, and systematically *overestimate* the error estimates on model parameters. In short, it provides a reasonably well defined method of making conservative error estimates.

The limits on the variations of a single parameter about a minimum $\chi_{min}^2$ in the $\chi^2$ distribution are determined by the variations in $\Delta\chi^2 = \chi^2 - \chi_{min}^2$, which is expected to follow a $\chi^2$ distribution with one degree of freedom. The formal 68% confidence interval is given by the range of values that produces $\Delta\chi^2 \leq 1$. In practice (see below) we find this criterion gives unrealistically small error estimates. A more conservative error estimate is found by renormalizing the $\chi^2$ of all the intermediate trial models in the database and collect all trial models with $\Delta\chi_r^2 \leq 15.1$ (the formal 99.99% confidence level). The error bars on the parameters are set to be the largest deviation from the best fit value. The error bars should not, however, be considered true 99.99% confidence level error bars because this assumes that the errors really are set by $N_{dof}$ independent, Gaussian-distributed errors. A more realistic assessment might be to consider them two standard deviation error bars. In practice, the correct way to estimate the significance of the errors would be with Monte Carlo simulations of the data, but Monte Carlo error estimation requires many realizations, each as time consuming as the original inversion, making it computationally impractical.

### 2.4. Systematic Effects of Using Clean Maps

The observed image obtained from a radio interferometer is $I_D = I \otimes B_D + N$ where $I$ is the true image of the source, $B_D$ is the response function of the interferometer or the dirty beam, and $N$ is the noise. Because of the irregular, discrete sampling of the u-v plane by the interferometer, the dirty beam has a complicated, slowly decreasing side lobe pattern that makes it difficult to interpret the dirty image $I_D$. This problem is overcome using non-linear image restoration methods such as Clean (Högbom 1974, Clark 1980, and Schwab 1984) or MEM (Cornwell & Evans 1985). The images we invert were generated using the Clean algorithm.

Any non-linear inversion method inevitably introduces some artifacts into the restored image. The restoring beam (or the entropy in MEM) eliminates short wavelength structure in the restored image and this finite resolution is not a problem in the LensClean restorations. There are difficulties, however, in how the methods restore the irregularly sampled regions of the Fourier plane. The fundamental problem is that the restorations of a lensed object are "local" restorations of the images that do not take into account the global constraint that the lensed images originate from a common source. LensClean works by applying this global constraint to determine the optimal lens model, so there are systematic errors associated with using a restored map rather than the raw visibilities. The systematic errors are intrinsic to using any reconstructed map instead of the raw visibilities, and a maximum entropy (MEM) or NNLS (Briggs et al. 1994) reconstruction also fills in the missing visibilities without accounting for the lens constraints. Unfortunately the size of the dirty beam and the need to do large numbers of LensCleans to survey the potential model make using the raw visibility data too computationally costly.

We can examine the systematic errors introduced by using the Clean map by restoring the visibilities using seven different Clean loop gains $\gamma_g$ to produce a sequence of 8 GHz images (see Table 2). Each restoration fits the measured visibilities "exactly", but differs in how the unmeasured visibilities are interpolated. The basic structure of the ring is preserved when we vary $\gamma_g$, but some of the small scale structures are not. Maps with larger $\gamma_g$ begin to develop stripes and clumps of Clean components. We started with the best fit model found by optimizing all seven parameters on the 8 GHz map with $\gamma_g = 0.1$ (the model is discussed in §3), and then we LensCleaned all seven maps using this fixed model to see how the residuals varied. Since the model was optimized on the $\gamma_g = 0.1$ map, this map has the smallest residuals. The largest residual is 17% larger and the average residual is 9% higher than the residual of the map with $\gamma_g = 0.1$. The origin of these differences is the change in the interpolated structure between the maps, and the magnitude of the residuals suggests that there is a systematic error due to the use of the Clean maps of about 20 $\mu$Jy per pixel rms in addition to the intrinsic noise of 35 $\mu$Jy per pixel.

Next we tested to see if the properties of the converged models are changed by the differences in the reconstructions, so we optimized the lens model for each of the seven maps (see Table 3). Reconverging the models halves the spread in the residuals found when we held the model parameters fixed. The converged models are all very similar. The scatter in the parameters amounts to 0.3% in the critical radius, 0.″003 in the lens position, 1.6% in the dimensionless shear, and 0.4° in the angle of the shear. These systematic errors are smaller than the statistical errors we derive in §3. These tests suggest that the results of the inversions are not dominated by the systematic errors from using reconstructed maps.

## 3. The 8 GHz Models

We focus our models on the 8 GHz maps because they have the best combination of resolution and signal to noise. The 8 GHz map has an intrinsic rms noise per pixel of 35 $\mu$Jy and an average beam FWHM of 0.″19 (see Table 1). This means that the number of resolution elements inside the tangential critical line is 65, so the number of degrees of freedom is $N_{dof} = 58$ for seven model parameters. We first examine the de Vaucouleurs model for the lens galaxy to establish a baseline. Then we discuss the $\alpha$ model qualitatively to understand the expected dependence of the goodness of fit on the core radius $s$ and exponent $\alpha$. Next we examine the numerical inversions for the $\alpha$ model. We also examine the effects of interpreting the central component D as radio emission from the lens galaxy instead of as a lensed image. Finally we consider the effects of two nearby galaxies seen in the infrared on the lens model.

### 3.1. The 8 GHz de Vaucouleurs Models

The de Vaucouleurs (1948) model has only one structural parameter, the effective radius $R_e$. For a fixed value of $R_e$ the value of $b$ is fixed by the average ring radius to high precision. Table 4 and Figure 2 summarize the de Vaucouleurs (1948) models as a function of effective radius. We find the same best fit effective radius whether we use $\chi^2_{tot}$, $\chi^2_{mult}$, or even just the peak residual. The best fitting model has $R_e = 0\farcs83 \pm 0\farcs13$, $\sigma_{tot} = 39.5$ $\mu$Jy, $\sigma_{mult} = 55.6$ $\mu$Jy and a peak residual of 342 $\mu$Jy. The total mass of the galaxy in the best fit model is $M = 1.17 h^{-1} [D_{ol} D_{os}/(2 r_h D_{ls})] \times 10^{12} M_\odot$. In an Einstein-DeSitter cosmology with source and lens redshifts of $z_s = 2.0$ and $z_l = 0.5$, the total mass is $M = 2.57 \times 10^{11} h^{-1} M_\odot$. Figure 3 shows the reconstructed image, the residuals, and the inferred source for the best fit model. Although the reconstructed image is very similar to the original image and the peak residuals are only 5% of the original peak of the map, there clearly are significant residuals. The best fit model has $\chi^2_{tot} = 825$ and $\chi^2_{mult} = 703$. These values for the $\chi^2$ statistics are formally 60 (!) standard deviations from the target value $N_{dof} = 57$. If we understand the $\chi^2$ statistics and we are not dominated by unknown systematic errors, then we can completely reject the de Vaucouleurs + external shear model for the lens. The best test of whether we do understand the errors is whether we can find models with significantly lower $\chi^2$ values.

The error bar associated with the effective radius is determined by the point where $\Delta\chi^2_r = 15.1$ in the renormalized $\chi^2_r$ statistic. This is an extremely conservative way of determining parameter errors. In $\chi^2$ parameter estimation, a one standard deviation change in the parameter corresponds to $\Delta\chi^2 = 1$ whereas our standard error bar is defined using $\Delta\chi^2_r = 15.1$. We can illustrate the conservatism of our parameter errors by noting that the error bar on $R_e$ using $\Delta\chi^2_r = 1$ in the rescaled statistic is $0\farcs05$, and the error bar using $\Delta\chi^2 = 1$ in the unrescaled statistic is $0\farcs01$, compared to our standard error estimate of $0\farcs13$. The large difference between the $\chi^2$ and $\chi^2_r$ statistics is caused by the large rescaling of the errors needed to make the de Vaucouleurs models a good fit to the data. Hammer et al. (1991) estimated $R_e \sim 0\farcs5$ in their optical data but they did not estimate their uncertainties. An effective radius of $R_e = 0\farcs5$ for the mass distribution is very strongly ruled out by the inversions, but the best fit value of $0\farcs83$ may be consistent with the optical data if the uncertainties in the optical estimates are large enough. Models of MG1654+134 (Kochanek 1994) also found that the best fit effective radius for the mass distribution was larger than the best fit to the optical data.

### 3.2. A Qualitative Picture of the $\alpha$ Model

Of the seven free parameters in the $\alpha$ model, we want to focus on the two parameters describing the shape of the mass distribution: the exponent $\alpha$, and the core radius $s$. The full LensClean models optimize the average fit over the entire map, but we can qualitatively understand the models by considering how the models simultaneously fit both the compact components and the overall ellipticity of the ring. The location of component D fixes the position of the lens reasonably accurately. We need to find three constraints to determine the parameters $b$, $\gamma$, and $\theta_\gamma$, and a fourth constraint to find a relation between the core radius $s$ and the exponent $\alpha$.

The two compact components must be images of the same source, so the relative positions and fluxes of the A and B components give us the three constraints needed to determine $b$, $\gamma$ and $\theta_\gamma$. We derive the positions and fluxes of the compact components from the 22 GHz map where they are best isolated from the extended ring (see Table 1). The 22 GHz flux of Component A is 2.86 mJy, and that of Component B is 3.85 mJy. The fourth constraint we impose is the requirement that the ellipticity of the critical line matches the measured ellipticity of the ring, $e_{ring} = 0.18 \pm 0.02$.

After fitting $b$, $\gamma$, and $\theta_\gamma$ to the three constraints needed to model the compact components, Figure 4a shows contours of the estimated ring ellipticity as a function of $\alpha$ and $s$. The ellipticity varies monotonically with $\alpha$ and $s$, and increasing values of the exponent or the core radius lead to smaller ellipticities. Models with $\alpha \gtrsim 0.9$ cannot simultaneously fit the compact components and produce a ring as elliptical as observed. Models with $\alpha \lesssim 0.9$ can fit both the compact components and the ring ellipticity. For each value of $\alpha$ there is a restricted range of core radii consistent with the constraints, and smaller values of $\alpha$ require larger core radii. The Kochanek et al. (1989) models of MG1131 had $\alpha = 1$ (albeit with a different elliptical structure) and produced rings that were more circular than the data, consistent with this qualitative picture.

The core radius also controls the flux of the central image; too large a core radius makes it too bright, and too small a core radius makes it too faint. The main contributor to the central lensed image in a ring system like MG1131 is the ring. This makes analytic models for the central image difficult because of the approximations needed to estimate the contribution from the ring. Let the central magnification of the lens be $M_0$ (eqn 2-10 for the $\alpha$ model). The compact component A (or B) with flux $F_A$ and magnification $M_A$ contributes flux $F_A M_0 M_A^{-1}$ to the flux of component D. If we fit the lens model to the compact components A and B, we get a good estimate of their contribution to component D. The contribution of the ring is difficult to calculate analytically because the only easily measured number is the total flux of the ring $F_{ring} = 100$ mJy at 8 GHz. The contribution of the ring to component D depends on the average of the number of images and the magnification for the extended emission, $\langle NM_{ring}\rangle$, which we estimate to be in the range $5 < \langle NM_{ring}\rangle < 20$. The predicted flux of component D is then

$$F_D^M = M_0 \left[ \frac{F_A}{M_A} + \frac{F_{ring}}{\langle NM_{ring}\rangle} \right] \tag{3-1}$$

compared to the observed flux of $F_D = 320$ $\mu$Jy at 8 GHz. Figure 4b shows where the $\alpha$ model fit to the compact components can reproduce the flux of component D for several values of $\langle NM_{ring}\rangle$. The core radius needed to fit the flux of component D is consistent with the core radius needed to fit the ring ellipticity only when $\alpha \lesssim 0.8$.

### 3.3. The 8 GHz $\alpha$ Models

We are primarily interested in the structural parameters $\alpha$ and $s$ that control the radial shape of the lens potential. The qualitative discussion in §3.2 suggests that only limited ranges of $\alpha$ and $s$ can successfully model the lens. Figure 5a shows an example of how the $\chi^2$ residual varies with the core radius for a fixed value of $\alpha = 0.6$. There is a well defined minimum of the $\chi^2$, which allows us to estimate the best fit value for the core radius and its error bars. Figure 5b shows how these values vary as a function of the exponent $\alpha$.

Table 5 summarizes the models derived from the 8 GHz image. None of the models reaches the formal target $\chi^2$ of $N_{dof} = 58$. The best fit model has $\chi^2_{tot} = 740$ and $\chi^2_{mult} = 598$. These $\chi^2$ values are significantly smaller than the best fit de Vaucouleurs (1948) model of $\chi^2_{tot} = 825$ and $\chi^2_{mult} = 703$. If we rescale the errors so that the best fit $\alpha$ model has $\chi^2_r = N_{dof}$ then the de Vaucouleurs (1948) models are 0.63 (0.94) standard deviations worse in the total (multiple image) error estimates. If we do not rescale the errors, the significance of the difference of 8 (10) standard deviations is much larger. This change in the significance of the differences shows the conservatism inherent in the rescaling process. The formal values of the $\chi^2_{mult}$ estimates are about 50 (!) standard deviations from the target of $N_{dof} = 58$

Such a poor fit to the data requires some justification. Part of the problem is the automated stopping criterion required by LensClean. The Cleaning procedure is stopped when a major cycle of the LensClean fails to reduce the peak residual in the map. This is a very good criterion for deciding when it is no longer profitable to pursue the current model. It does, however, mean that the procedure stops before the mean square residuals used in the $\chi^2$ statistics are truly minimized. If we take the best fit model and use 30,000 components, instead of the 1500 used with the standard stopping criterion, the $\chi^2$ estimates drop to $\chi^2_{tot} = 336$ and $\chi^2_{mult} = 315$. For the de Vaucouleurs model using 30,000 components reduces the $\chi^2$ estimates to $\chi^2_{tot} = 430$ and $\chi^2_{mult} = 407$. Both $\chi^2$ values of the best de Vaucouleurs model are still considerably larger than the ones from the best $\alpha$ model, but the differences between the two models are 1.5 (1.7) standard deviations in the rescaled total (multiple image) error estimates and 9 (9) standard deviations in the unrescaled total (multiple image) error estimates. Thus a different stopping criterion reduces the absolute residuals, but leaves the differences between models unaffected. We also know from our study of the systematic variations in the errors of §2.4 that there is some additional contribution to the noise from systematic errors. The analysis of §2.4 suggests adding 20 $\mu$Jy per pixel (in quadrature) to the noise estimates to encompass the systematic variations. This would reduce $\chi^2_{tot}$ and $\chi^2_{mult}$ to 253 and 237 respectively, which is still much larger than required for a good fit to the data. Comparable models for the radio ring MG1654+134 (Langston et al. 1989) approach the target values for the $\chi^2$ much more closely (Kochanek 1994). This suggests that in the final analysis our best fitting model is not totally consistent with the data, and in §5 we will discuss the implications of this conclusion. For the moment we will discuss the differences between models using the renormalized estimate of the $\chi^2$.

Figure 6 shows the two renormalized $\chi^2$ statistics as a function of the exponent $\alpha$ after optimizing the core radius for each value of the exponent. With the renormalization, we find that models with $0.4 \lesssim \alpha \lesssim 0.8$ are within $\Delta\chi^2_r = 15.1$ of the minimum. We should again note the conservatism of this range estimate: using normal one standard deviation errors with $\Delta\chi^2 = 1$ on the rescaled (unrescaled) values of the $\chi^2$ gives errors on the value of $\alpha$ of $\pm 0.06$ ($\pm 0.02$). The peak residual in the best fit map is 357 $\mu$Jy, and all the acceptable models ($0.4 \lesssim \alpha \lesssim 0.8$) have peak residuals smaller than 421 $\mu$Jy/pixel. The isothermal model ($\alpha = 1$) lies outside the permitted range, making this the first example of a lens whose radial mass distribution apparently cannot be modeled by a quasi-isothermal potential. Figure 7 shows the residual map, the inferred source that produces the lens, and the reconstructed image for the best fit model given in Table 5. The source is convolved with a "proper" beam corrected for the magnification (see §3.4 in Kochanek & Narayan 1992). The reconstruction is a plausible extragalactic radio source consisting of a bright core, two radio lobes and possibly a short jet. Since the residuals are relatively small, the reconstructed image and the observations cannot easily be distinguished by eye. The reason the global $\chi^2$ is large is clear from the many contours above the noise in the residual map.

Table 6 and Figure 8 summarize some of the physical properties of the models such as the mean ring radius, the ring ellipticity, the mass interior to the ring, and the time delay between the compact components. The lens position is fixed and independent of the other lens parameters because the symmetry of the ring and the position of component D leaves little room for shifts in the lens position. Similarly, the position angle of the shear is model independent because the tangential critical line of the lens must have its major axis at right angles to the major axis of the ring as discussed by Kochanek et al. (1989). Models with $0.4 \lesssim \alpha \lesssim 0.8$ give roughly the same ellipticity ($\sim 0.17$) which matches the value measured from the ring ($0.18 \pm 0.02$). Figure 4 superimposes the best fit core radii and their error bars as a function of $\alpha$ on the analytic estimates made in §3.2. As expected, the good models track the band of ellipticities consistent with the ellipticity of the ring and the flux of the central image. Note that the best fit isothermal model derived from fitting the ring gives an estimated ring ellipticity of 0.194, while the analytic estimate derived from fitting the compact components in §3.2 predicted that an isothermal model fit to the compact components would give an ellipticity of 0.14. In the inversion the dominant constraint is the structure of the ring, so the isothermal models make elliptical rings at the price of not fitting the compact components as accurately. For comparison, the Kochanek et al. (1989) models fit a 15 GHz map of MG1131 in which the dominant constraints were the compact components and they produced rings that were too circular. Figure 8 presents plots of the mass profile and the deflection angle as a function of projected radius. As one would expect, the mass interior to the ring is accurately determined and is insensitive to the model adopted, and all models give the same mass and deflection values at the average ring radius. The predicted time delay varies slightly from one model to the other, with the largest variation of $\pm 9\%$ between models with $0.4 < \alpha < 0.8$. The formal one standard deviation errors on the time delay are $\pm 1\%$ for $\Delta\chi^2 = 1$ and $\pm 4\%$ for $\Delta\chi_r^2 = 1$.

### 3.4. Do the Results Depend on the Interpretation of Component D?

The core radius plays a variety of roles in the lens models. It controls the flux of the central component D, the size of the multiply image region and the fraction of component C that is multiply imaged, and (when $\alpha < 1$) it adjusts the slope of the lens deflection near the critical line. In this section we examine whether the core radius is determined by the flux of the central image or by the need to correctly fit the ring. We do this by subtracting component D from the map and then redoing the models. The central component D is probably a lensed image, but it could also be emission by the lens galaxy (Chen & Hewitt 1993). By looking at models with D subtracted, we also check the effects of misinterpreting the source of D.

We modeled the modified map at a fixed value of $\alpha = 0.6$ to redetermine the limits on the core radius. Since the main purpose of this experiment is to understand the other constraints on the core radius, examining a single value of $\alpha$ should be sufficient. Figure 9 shows that the residuals as a function of core radius are similar to the results without component D. This demonstrates three important points. First, the core radius of the potential is largely constrained by the location of the caustics and the geometry of the ring, rather than by the flux of the central component. This means that simply fitting the core radius to produce the flux of the central component can produce a qualitatively and quantitatively worse model than fitting the core radius to get the best average fit to the extended structure. Second, any misinterpretation of component D does not significantly bias the results of our current models, because the flux density of D is not a major contributor to the constraints. Third, because the model must have a substantial core radius to fit the larger lensed structures, there must be a central lensed image. In the best 8 GHz model we predict a flux density for component D of 266 $\mu$Jy compared to the measured flux density of $320 \pm 60$ $\mu$Jy, so most, if not all, of the flux of D is a lensed image and not emission from the lensing galaxy.

### 3.5. The Effect of Two Nearby Galaxies

Annis (1992) and Larkin et al. (1994) note the presence of two fainter galaxies near the MG1131 system. Annis (1992) named the galaxies C and D, but we rename them G1 and G2, respectively, to prevent confusion with the radio components in the system. We will call the primary lens galaxy L. The two galaxies are both approximately 3 arcseconds from the lens position. Relative to the best fit lens position G1 is at (-0.″09, -3.″15) and G2 is at (2.″04, 1.″89). The $K$ magnitudes of L, G1, and G2 are 16.8, 20.8 and 21.2 respectively (Larkin et al. 1994). These two galaxies are so close to the ring that their contribution to the lensing may be important. We need to investigate whether our parameterization of the lens models is adequate to model their effects and whether they may account for the unusually large ellipticity of the ring.

Assuming G1 and G2 are at the same redshift as L, the magnitude differences and the Faber-Jackson (1976) relation predict the ratios of the critical radii of G1/G2 and L to be $b_{G1}/b_L = 0.16$ and $b_{G2}/b_L = 0.13$. This means that G1 and G2 have a perturbative effect on the lens model, and we can expand the potentials of G1 and G2 as a power series centered on the primary lens galaxy L. The expansion of the perturbing potential of each galaxy to second order in $b_G/r_G$ is

$$\phi_e = \text{constant} + a_1 x + a_2 y + \frac{1}{2}\kappa_e r^2 + \frac{1}{2}\gamma_e r^2 \cos[2(\theta - \theta_e)]. \qquad (3\text{-}2)$$

The constant and linear ($a_1 x + a_2 y$) terms have no effects on the model, so the first terms that modify the lens model are the convergence $\kappa_e = \Sigma/\Sigma_c$ produced by the extra surface density of G1 and G2 near the ring, and the shear $\gamma_e$ in their gravitational field (Alcock & Anderson 1985, 1986; Falco, Gorenstein & Shapiro 1988) If the perturbing galaxies have the same monopole structure as the lens galaxy and a small core radius ($s/r_G \ll 1$), then the convergence and shear from one of the two galaxies are

$$\kappa_G = \frac{\alpha_G}{2}\left(\frac{b_G}{r_G}\right)^{2-\alpha_G} \qquad \text{and} \qquad \gamma_G = \left(1 - \frac{\alpha_G}{2}\right)\left(\frac{b_G}{r_G}\right)^{2-\alpha_G} \qquad (3\text{-}3)$$

where $\alpha_G$ and $b_G$ are the monopole lens parameters associated with the external galaxy, and $r_G$ is the distance from the center of the principle lensing galaxy to the center of the external galaxy.

These two lowest order terms have no effect on our models. The convergence term ($\kappa_e$) has no observable consequences in a lens model, it simply rescales the parameters of the models (Alcock & Anderson 1985, 1986, Falco, Gorenstein & Shapiro 1988). The convergence term means that the true critical line of the lens galaxy $b_L = (1 - \kappa_e)b$ where $b$ is the parameter fitted in the lens models. The convergence produced by G1 and G2 at the ring is small; even for the slowly decreasing density of the isothermal model $\kappa_e \simeq 0.05$. Thus the convergence from the nearby galaxies introduces only a small rescaling of the lens parameters.

The shear produced by the two galaxies is identical to the external shear we use in the models. The superposition of three separate external shears for the main lens galaxy and the two perturbing galaxies is simply an external shear model with a different ellipticity and position angle. Thus our quadrupole structure is in many ways a better model for the effects of the perturbing galaxies than for the primary lens. The orientation of the perturbing shear, which depends only on the mass ratio and relative positions of G1 and G2, has $\theta_e = -23°$. This is only 3° from the position angle of the shear in our lens models - a remarkable coincidence! The strength of the shear depends somewhat on the form of the monopole, with $\gamma_e = 0.030$ for an isothermal model and $\gamma_e = 0.003$ for a point mass model. Both values are small compared to the total shear of $\gamma \simeq 0.13$ needed to fit the ring in our models. Therefore, unless our estimate of their masses relative to L is in error, G1 and G2 do not produce the shear required to model the ring in MG1131. If we continue a power expansion of the effects of the perturbing galaxies, the next order terms are smaller by another factor of $b_G/r_G \sim 0.05$ and they will change the deflections of rays by at most 0.″002. This is smaller than the changes in the deflections of rays produced by our typical errors on parameters, so it is probably an unmeasurable perturbation in our current models.

The coincidence between the modeled position angle of the external shear and that computed for G1 and G2 prompts us to explore scenarios in which the shear can be dominated by the external galaxies. Any such scenario would require increasing the influence of G1 and G2 relative to the principle lensing galaxy. In Figure 10 we show the effects on the lens critical lines and the inferred shear at the ring of reducing the luminosity difference between G1/G2 and L by factors of 25, and 100. We do this for both point mass and isothermal lens models, rescaling the parameters of the primary lens galaxy for the convergence introduced by G1 and G2. We see that the observed K band luminosity ratios must be in error by a factor of 25 for the isothermal models before the shear from G1 and G2 is large enough to match that required by the models. More concentrated models require still larger changes from the observed luminosity ratio, reaching a factor of 100 for the point mass models.

If the Faber-Jackson (1976) relation holds, then changing the relative lensing strength of G1/G2 and L must come either from reducing the luminosity of L or increasing the intrinsic luminosity of G1 and G2. The luminosity of L is inferred by subtracting the contributions from the lensed images of the compact radio components A and B. The optical residual is lumpy and ill-formed (Larkin et al. 1994) and Hammer et al. (1991) claim to see the ring in which case more of the flux of L might be from the source rather than the lens. Nonetheless, it is hard to see how the estimates could be off by a factor of 25 in luminosity. The infrared images of Larkin et al. (1994) also indicate that the MG1131 lens galaxy is dusty. If extinction by L is dimming the light from G1 and G2, the amount of dust required is large, and preserving the position angle of the external shear would require the unlikely coincidence that G1 and G2 have the same extinction. However, the position angle is not a sensitive function of the G1/G2 mass ratio, and it would differ by less than $10^\circ$ for magnitude differences as large as 1.8. Finally, if G1/G2 were at a different redshift than L, then the estimates would also be incorrect. It is very hard to evaluate this possibility because of the complete absence of redshift information on L, G1, G2, and the source. A surrounding cluster could easily generate the required shear, but if the shear is generated by a cluster there is no reason for it to be aligned with the predicted shear from G1 and G2.

## 4. The 5 GHz and 15 GHz $\alpha$ Models

Gravitational lenses are achromatic so the models that fit the 8 GHz data should also fit the 5 GHz and 15 GHz data. Because the spectral index of the data is not constant the different frequencies emphasize different parts of the image. For example, higher frequencies emphasize the compact components over the ring because the compact components have flatter spectral indices than the extended ring. We model the 5 and 15 GHz maps using the same procedure we used for the 8 GHz map, but we limit the analysis to the range $\alpha$ ($0.4 < \alpha < 0.8$) and $s$ (see figure 5(b)) that produce acceptable fits to the 8 GHz data. The target $\chi^2$ values for the 5 and 15 GHz maps are 15 and 153 respectively, different from that at 8 GHz because of the different restoring beams. We explore only the $\alpha$ models because they have the greatest potential for producing a different result. The two questions we examine are whether the same model produces the best fit at the other two frequencies, and whether the model is a good fit at the other two frequencies.

Table 7 summarizes the converged models for the 5 GHz map, and Figure 11a shows the renormalized $\chi^2$ as a function of $\alpha$. The dashed line, calculated from the formalism outlined in section 2.4, shows the limit where $\Delta\chi^2_r = 15.1$. The best fit model has $\alpha = 0.5 \pm 0.3$, and the results of the models $0.4 \lesssim \alpha \lesssim 0.8$ are comparable to the the 8 GHz results. The absolute values of the $\chi^2$ statistics - 407 (363) in the total (multiply) image region - are again much larger than the target value of $N_{dof} = 15$ by 71 (63) standard deviations. The peak residuals in the map are only $< 3.8\%$ of the peak brightness; however, because the 5 GHz map has a large signal-to-noise ratio this is still considerably larger than expected. Figure 12 shows the residual map, the inferred source, and the reconstructed image for the best fit 5 GHz $\alpha$ model. We again find a model that to the eye looks perfectly acceptable, and can be formally rejected by the goodness-of-fit criteria.

All converged lens parameters are consistent with the ones in the 8 GHz models given the uncertainties. The lens position is systematically shifted by $0.''02$ in the $x$ coordinate and $0.''01$ in the $y$ coordinate. The shifts are probably due to systematic problems in the registration of the two images because there are no strong compact sources in the map that could be used as position standards. The 5 and 8 GHz observations were taken on different dates with different phase calibrators leading to positional uncertainties on that order (see Chen & Hewitt 1993). The systematic uncertainties in registering the different maps are larger than the uncertainties in the lens positions from the inversions of the independent maps. This highlights the difficulty of attempting simultaneous models of several different maps using LensClean (Kochanek & Narayan 1992).

Table 8 summarizes the converged models for the 15 GHz map, Figure 11b shows the residuals as a function of $\alpha$, and Figure 13 shows the residual map, the inferred source, and the reconstructed image of MG1131 at this frequency. Once again we find that the lens parameters are consistent with those found in the 8 GHz modeling. The absolute values of the $\chi^2$ statistic - 1047 (427) in the total (multiply) image region - are again much higher than the target value of $N_{dof} = 153$.

## 5. Conclusions

The most fundamental conclusion of this paper is that none of the models is statistically consistent with the data. The $\chi^2$ estimates for the best fit models are typically 50 standard deviations from the expected $\chi^2 = N_{dof}$. Some of the discrepancy arises because we use a finite number of Clean components. Using a Clean map does not appear to be an important source of systematic errors. When we try to correct the $\chi^2$ statistics for the finite number of components and the systematic errors, we can halve the value of the $\chi^2$. The statistical differences between models were not affected by the corrections, and reducing the problem to being 25 standard deviations from the expected value of $\chi^2$ still means the models are unacceptable. LensClean models of the radio ring MG1654+134 (Kochanek 1994) with comparable resolution, flux, and noise were statistically consistent with the data, so it is unlikely that the problem is intrinsic to the method. We surmise that the problem is caused by the simple angular structure of our lens models.

Nonetheless, we find clear, best fitting models for the 5, 8, and 15 GHz maps given these limitations, with peak residuals smaller than 5% of the original peak of the map and reconstructions that are visibly almost identical to the data and small uncertainties in the converged parameters even after compensating for the statistical problems. The best fitting model is the $\alpha$ model, which has a monopole potential with $\phi \propto (r^2 + s^2)^{\alpha/2}$. The best models have $\alpha = 0.6 \pm 0.2$ and surface density profiles that asymptotically decline as $r^{-1.4\pm0.2}$. Assuming that the external shear well represents the angular structure of the lensing potential, we find that an isothermal lens ($\alpha = 1$) with a surface density profile that asymptotically declines as $r^{-2}$ is inconsistent with the data - the first example in which a lens cannot be modeled by the isothermal profile. The isothermal models fail to fit the lens because they cannot simultaneously fit the constraints from the compact components and the ring given the fixed quadrupole structure of the lens.

We also fit de Vaucouleurs (1948) models to the 8 GHz image. We found that the best fit effective radius was $R_e = 0\farcs83 \pm 0\farcs13$, larger than the $R_e = 0\farcs5$ estimated from optical images of the lens. New optical data on the morphology of the image, with estimates of the uncertainties on fitted parameters like the effective radius would allow more quantitative comparisons of the fitted mass model to the optical properties of the lens. The best fit $\alpha$ models are considerably better than the best de Vaucouleur models. The two models differ by 10 standard deviations in the $\chi^2$, or one standard deviation in the rescaled $\chi_r^2$.

In the $\alpha$ models the core radius of the lens is a strong function of the asymptotic exponent $\alpha$. Models with steeply declining density distributions $\alpha \lesssim 1$ require finite core radii that are a reasonable fraction of the ring radius, while near isothermal models with $\alpha \gtrsim 1$ require small or zero core radii. The dominant constraint on the core radius is not, in fact, the central image in the lens but the need to fit the extended structure of the ring by properly locating the multiply imaged region, and fitting the slope of the deflection profile at the ring radius. While the central image is a visually appealing source of constraints on the model, models that set the core radius simply by fitting the central image can catastrophically fail to fit the ring. In an experiment fitting the data with the central component D subtracted we found the same estimate for the best fit core radius as when component D was present. In the best fit model the flux of the central component of the model is within one standard deviation of the measured flux of component D. Within the limits imposed by our overall goodness of fit problem, we can rule out the hypothesis that component D is emission by the lens galaxy.

The parameters of the best fit $\alpha$ model track the analytic requirements that the lens is centered on component D, produces the average ring radius, fits the positions and fluxes of components A and B as images of a common source and reproduces the observed ring ellipticity. Figure 4 shows that the best fit $\alpha$ models track the simple analytic models. These analytic requirements mean that the model has almost no freedom to fit the radial structure of the ring, and that the restrictive assumptions about the structure of the quadrupole of the lens are determining the radial structure of the lens. This suggests that a different angular structure, or an angular structure with more degrees of freedom, might break the forced coupling between radial structure and ring ellipticity. The large $\chi^2$ value for the best fit model means that there is still great room for improvement in the models. The models we have tried assume that the ellipticity of the ring is generated solely by the matter well outside the ring, and the surface densities of the lens are circularly symmetric. Since the apparent isophotes of elliptical galaxies exhibit ellipse-like features, it is very likely that potentials with circular symmetric surface densities are not sufficient in representing the true potentials of the lens galaxy. We suspect that a lens with true ellipsoidal isodensities plus an external shear would be the best model to try next. The MG1131 ring is very elliptical for a lens, so the higher order terms in the angular structure produced by a real ellipsoidal lens may be important in fitting the ring structure. It is possible that an isothermal radial distribution will be consistent with the data if the angular structure of the model is changed.

The two galaxies G1 and G2 about $3''$ from the lens perturb the lens model, primarily by adding an additional external shear. The external shear produced by the two galaxies is only 3° from the axis of the external shear fitted in the model. This is a remarkable coincidence. The magnitude of the shear produced by G1 and G2 is almost an order of magnitude too small to produce the ring even when the galaxies are modeled as extended mass distributions like the isothermal sphere. We estimate that the mass/velocity ratios estimated from the luminosities of the lens galaxy, G1, and G2 must be in error by a factor of 25 or more before G1 and G2 have enough mass relative to the primary lens galaxy to produce the observed ellipticity. There appears to be no plausible scenario to make the errors that large, so the alignment seems to be only a remarkable coincidence.

We measure several physical properties of the lens galaxy with very high accuracy in "dimensionless" or "angular" units. We cannot convert them into physical units because the redshifts of the lens and the source are still unknown. For example, the mass interior to the ring is determined to within 2% by the average ring radius. The compact components in MG1131 can be used to measure the Hubble parameter, $H_0$, if they are variable and we can determine a time delay. The accuracy of measurement of $H_0$ depends on the accuracies of both the time delay measurements and the lens modeling. Although we cannot determine the actual value of the time delay without knowing the redshifts (for a lens at $z_l = 0.5$ and a source at $z_s = 2.0$ in a Einstein-DeSitter cosmology the delay is approximately $58h^{-1}$ days, see Table 6), we can estimate the uncertainties in the time delay due to the models. The formal one standard deviation uncertainty in the time delay is $\pm 1\%$ for $\Delta\chi^2 = 1$ or $\pm 4\%$ for $\Delta\chi_r^2 = 1$. We estimate a maximum uncertainty of $\pm 9\%$ from the limits when $\Delta\chi_r^2 = 15.1$. The key observational problem in MG1131 is measuring the redshifts of the lens and the source.

The extended structure in MG1131 provides extremely tight constraints on the lens models, and there is still no model that is completely consistent with the MG1131 radio surface brightness. We can address whether there are systematic problems in the models by fitting the dirty maps of the lens to avoid the systematic errors associated with using the processed Clean maps, and by doing Monte Carlo simulations of the observations and the inversions to better understand the statistical and systematic errors. Both of the approaches are computationally very difficult because of time limitations. By studying the polarization maps we can further constrain or differentiate between different lens models. We know, however, that there is significant Faraday rotation in MG1131 (Chen & Hewitt 1993) and we cannot use the polarization data for lens models unless the Faraday rotation is removed. If we assumed that we knew the correct lens model or a reasonable approximation of it, we could use the lens model to separate the Faraday rotation intrinsic to the source from the Faraday rotation by the lens galaxy. The lens models need a more sophisticated angular structure than the external shear model. The next plausible model is an ellipsoidal primary lens galaxy with an external shear from the two perturbing galaxies G1 and G2.

**Acknowledgements:** We thank S. Conner and J. Ellithorpe for helpful discussions, and J. Larkin and C. Lawrence for providing unpublished data. CSK acknowledges the support of the Alfred P. Sloan Foundation and NSF grant AST-9401722. JNH and GHC acknowledge the support of a David and Lucile Packard Fellowship in Science and Engineering and a National Science Foundation Presidential Young Investigator Award.

Table 1. junk

| junk |
| --- |
| 0 |

Table 2. The First Gain Experiment

| gain | 0.20 | 0.15 | 0.10 | 0.08 | 0.06 | 0.05 | 0.02 | mean |
|---|---|---|---|---|---|---|---|---|
| $\sigma^{a}(\mu\mathrm{Jy/pixel})$ | 41.7 | 42.7 | 37.6 | 41.1 | 38.8 | 40.1 | 44.1 | $40.9 \pm 2.2$ |
| $\Delta\sigma^{b}(\mu\mathrm{Jy})$ | 362 | 382 | 343 | 404 | 363 | 376 | 416 | $378 \pm 25$ |

[a]The rms flux in the residual map.

[b]The peak flux in the residual map.

Table 3. junk

| junk |
| --- |
| 0 |

Table 4. junk

| junk |
| --- |
| 0 |

Table 5. junk

| junk |
| --- |
| 0 |

Table 6. Physical Properties of the 8 GHz $\alpha$ models

| $\alpha$ | $\langle r \rangle$[a] | $e$[b] | $M(\langle r \rangle)$[c] | $\Delta\tau$[c] |
|---|---|---|---|---|
| | $''$ | | $h^{-1} 10^{11} M_\odot$ | $h^{-1}$ years |
| 0.3 | 0.914 | 0.159 | 1.36 | 0.183 |
| 0.4 | 0.907 | 0.170 | 1.34 | 0.175 |
| 0.5 | 0.909 | 0.169 | 1.35 | 0.171 |
| 0.6 | 0.912 | 0.170 | 1.35 | 0.163 |
| 0.7 | 0.917 | 0.178 | 1.37 | 0.154 |
| 0.8 | 0.916 | 0.173 | 1.37 | 0.147 |
| 0.9 | 0.922 | 0.184 | 1.38 | 0.135 |
| 1.0 | 0.929 | 0.194 | 1.40 | 0.124 |
| $0.63 \pm 0.23$ | 0.914 | 0.172 | 1.36 | 0.159 |

[a]The average ring radius.

[b]The ellipticity of the ring.

[c]The masses and the time delays in the table are calculated assuming that $z_s = 2.0$ and $z_l = 0.5$ in an Einstein de Sitter universe. To correct for other combination of redshifts, multiply the mass by by $4.55 D_{ol} D_{os}/2 r_h D_{ls}$ and the $\Delta\tau$ by $3.03(1+z_l) D_{ol} D_{os}/2 r_h D_{ls}$.

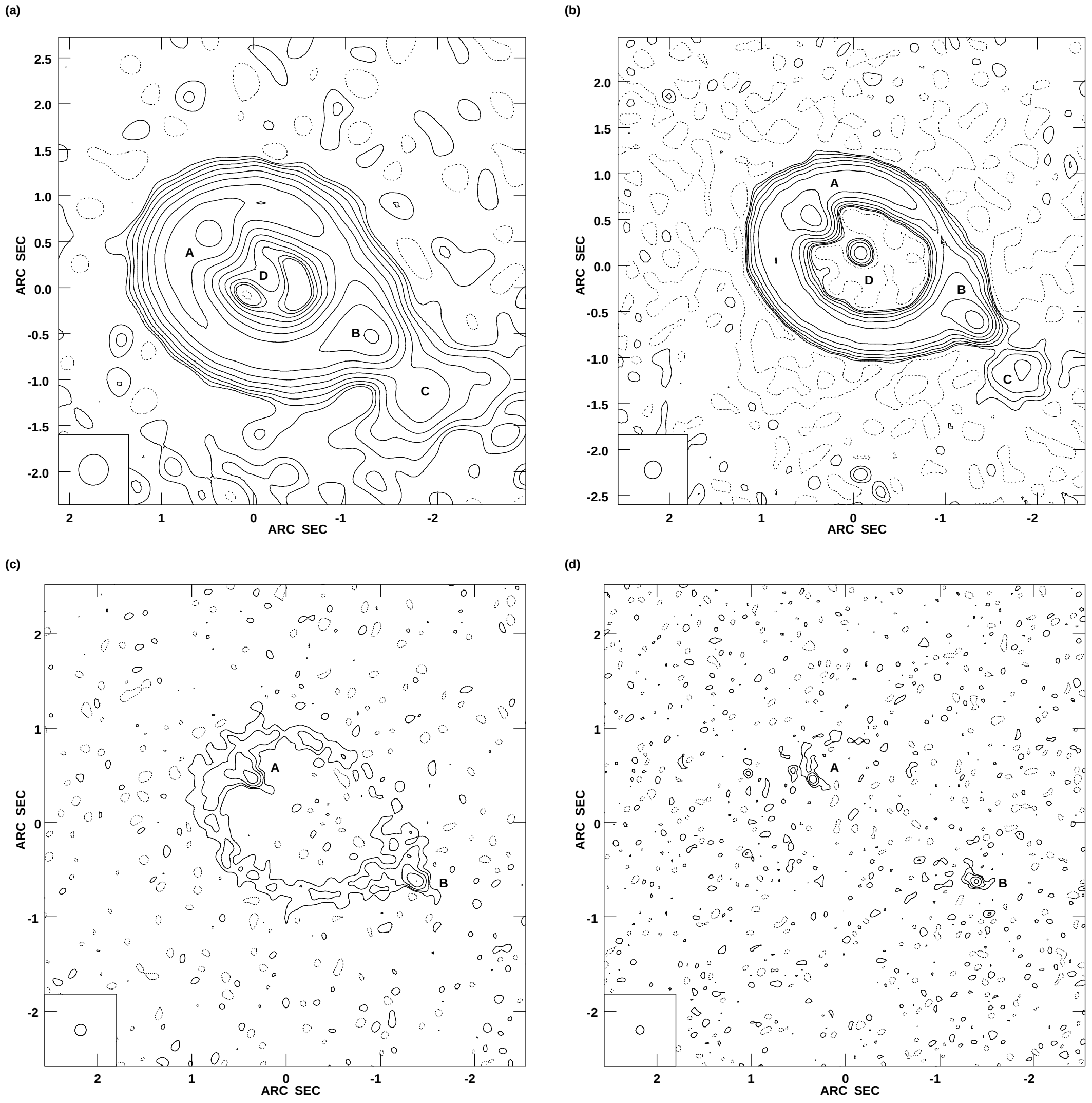

Fig. 1.— The total intensity VLA maps of MG1131+0456 at 5(a), 8(b), 15(c), and 22(d) GHz. In the 5 and 8 GHz maps, the contour levels are -1, 1, 2, 4, 8, 16, 32, 64, 128, 256 times the estimated noise in the maps (60 and 35 $\mu$Jy/pixel). In the 15 GHz map, the contour levels are -2, 2, 4, 8, 16, 32 times the estimated thermal noise in the map (130 $\mu$Jy/pixel). In the 22 GHz map, the contour levels are -4, 4, 8, 16 times the estimated thermal noise in the map (200 $\mu$Jy). The synthesized beam is shown in the lower left corner of each map.

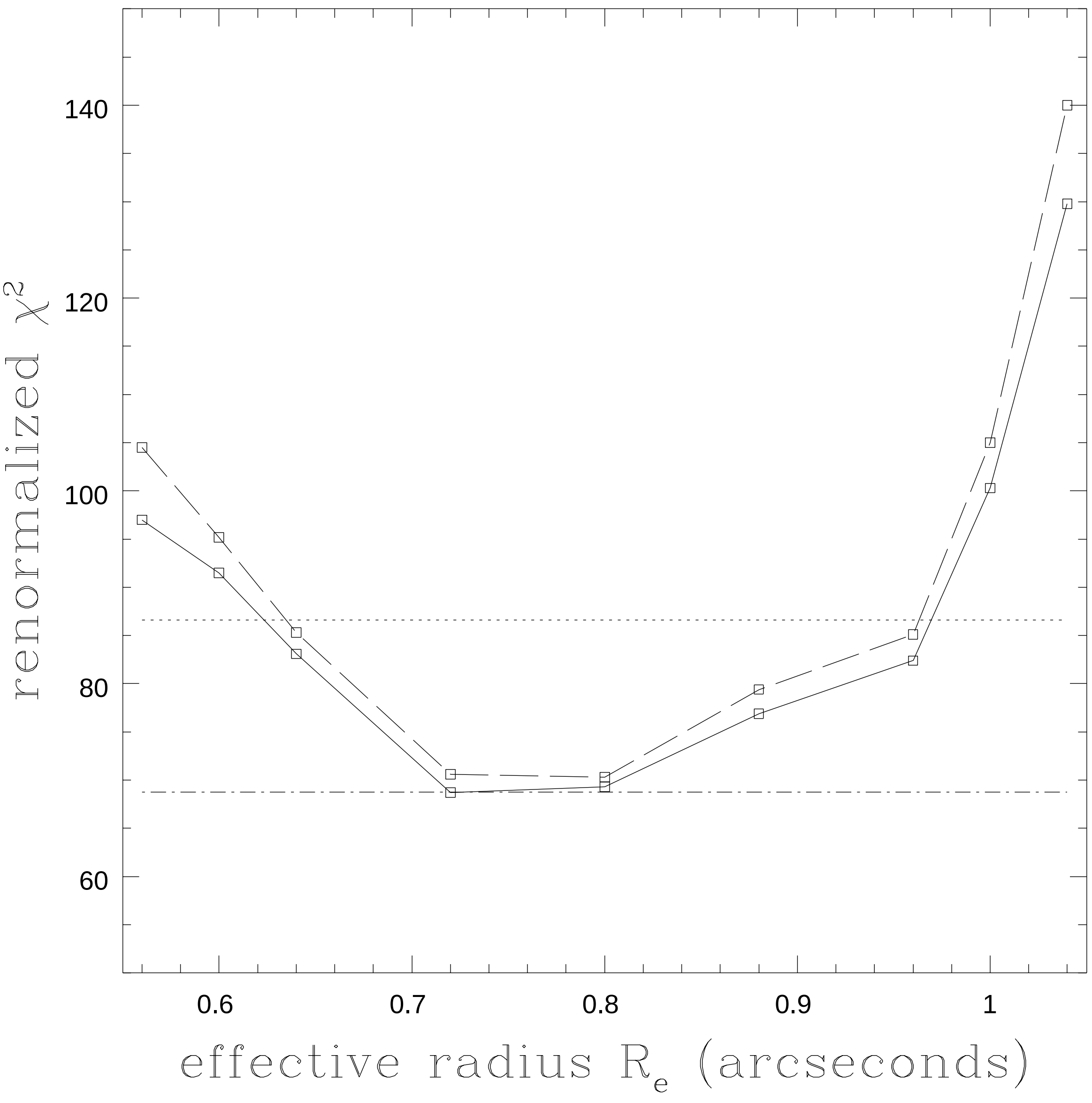

Fig. 2.— The renormalized $\chi^2$ of the 8 GHz de Vaucouleurs models as a function of the effective radius, $R_e$. To ease the comparison between the de Vaucouleurs models and the $\alpha$ models, the renormalization constant used here is the same as the one used in the 8 GHz $\alpha$ models. The dotted-dashed line represents the one standard deviation increase in $\chi_r^2$ from the best $\alpha$ model. The solid line shows the renormalized $\chi_{tot}^2$, and the dashed line shows the renormalized $\chi_{mult}^2$ of the de Vaucouleurs models. The dotted line shows $\Delta\chi_r^2 = 15.1$ from the best de Vaucouleurs model in the renormalized $\chi^2$ statistics.

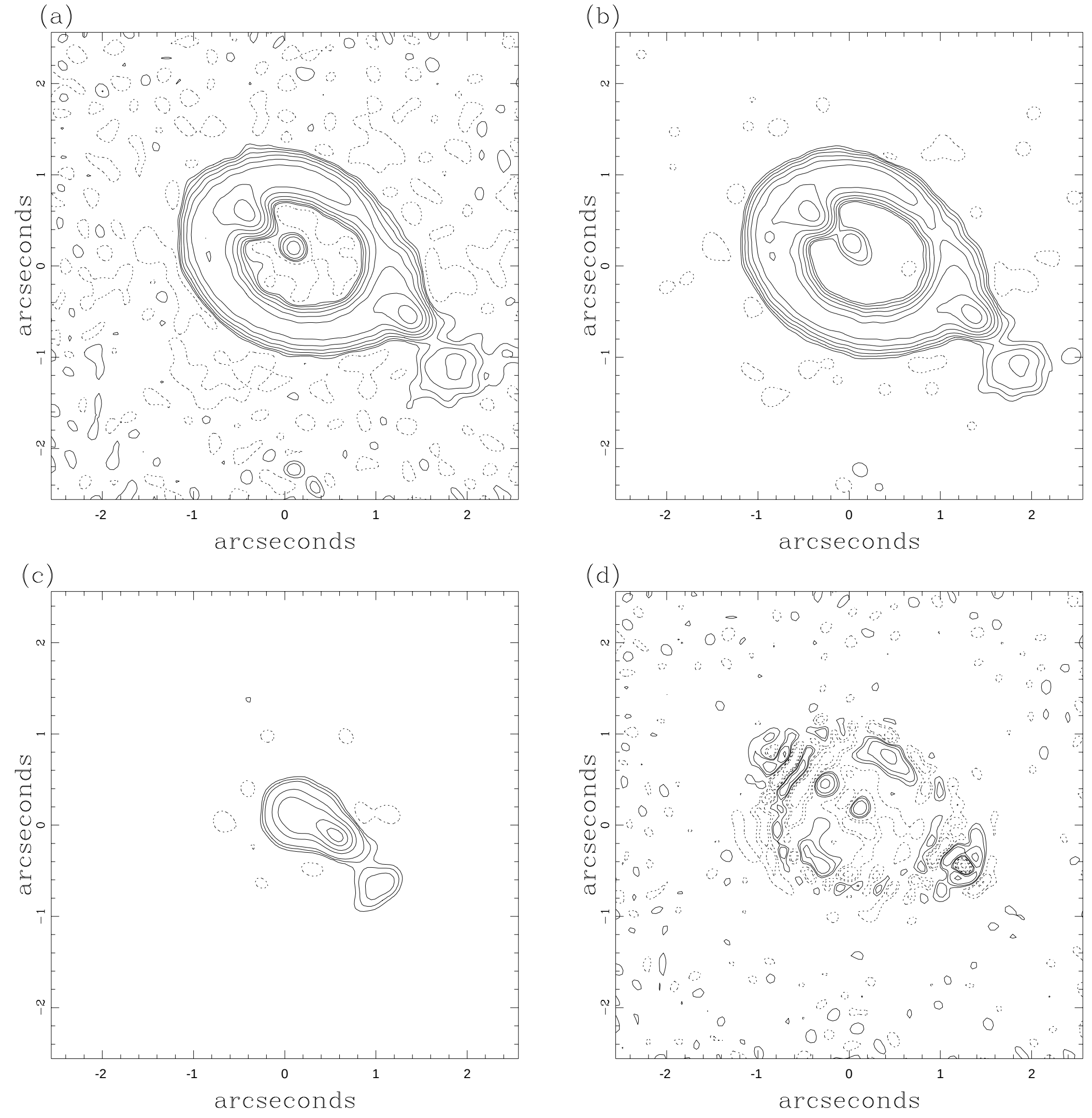

Fig. 3.— (a) The observed image, (b) the reconstructed image (c) the inferred source, and (d) the residual map obtained from the best 8 GHz de Vaucouleurs model. The contour levels in the figures are: (a) and (b) $-1, 1, 2, 4, 8, 16, 32, 64, 128, 256 \times 35\mu$Jy, (c) $-1, 1, 2, 4, 8, 16, 32, 64, 95\% \times 155\mu$Jy, and (d) $-8, -4, -2, -1, 1, 2, 4, 8 \times 35\mu$Jy. In a perfect reconstruction (a) and (b) would be identical.

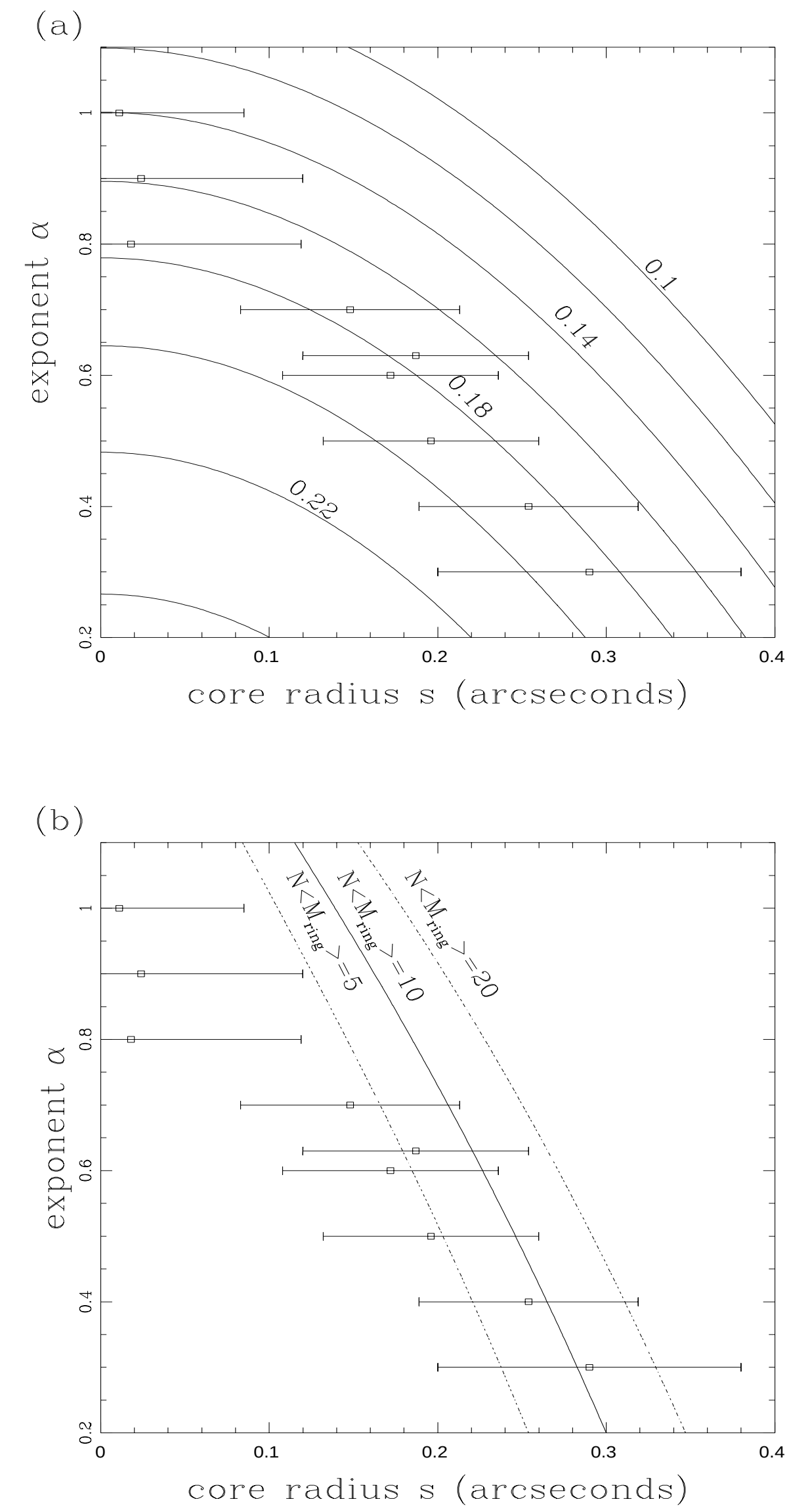

Fig. 4.— (a) A plot of the predicted ellipticity of the ring as a function of $\alpha$ and $s$. The contour levels show the predicted ellipticity from 0.1 to 0.24 with increment of 0.02. The measured ellipticity of the ring is $0.18 \pm 0.02$. The data points display the best fit core radii and their error bars for the 8 GHz $\alpha$ models as a function of $\alpha$. The best fit core radii of the acceptable models overlap the predicted ellipticity. (b) The estimated core radius as a function of $\alpha$ estimated from the flux density of D (320 $\mu$Jy), the magnification of the compact components, and the flux density of the ring (100 mJy) using three possible values of $N\langle M_{ring}\rangle$ (5, 10, and 20). The best fit core radii and their error bars for the 8 GHz $\alpha$ models are also shown (the data points).

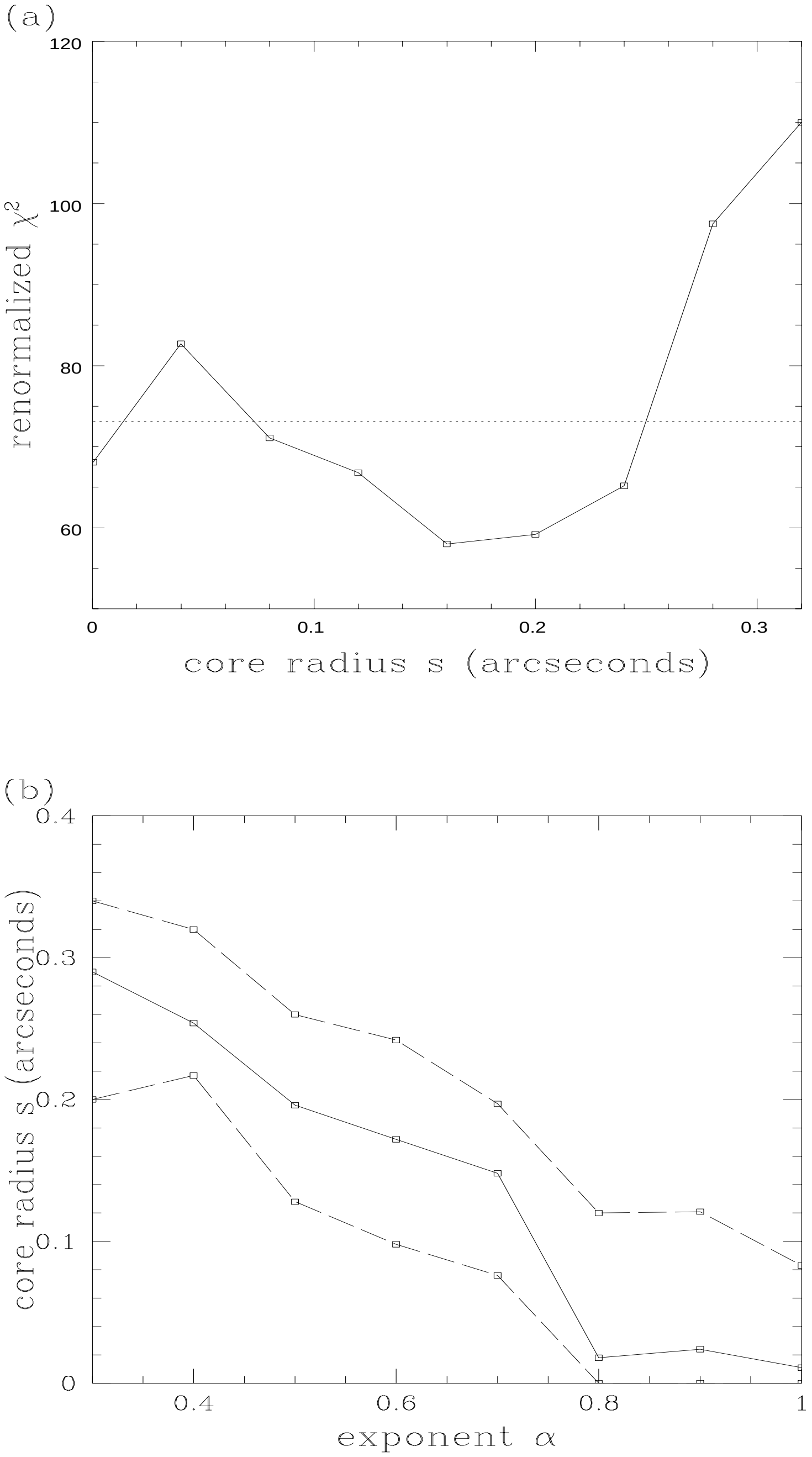

Fig. 5.— (a) The renormalized $\chi^2$ as function of core radius $s$ for $\alpha = 0.6$. This figure demonstrates how the limits on $s$ at a fixed value of $\alpha$ are determined. The dashed line shows $\Delta\chi^2 = 15.1$ in the renormalized $\chi^2$ statistics, and the range of the core radii under the dashed line determines the limits on $s$. The method is applied to every value of $\alpha$ to determine the limits on $s$. (b) The optimized value of the core radius $s$ (solid) and its limits (dashed) as function of $\alpha$.

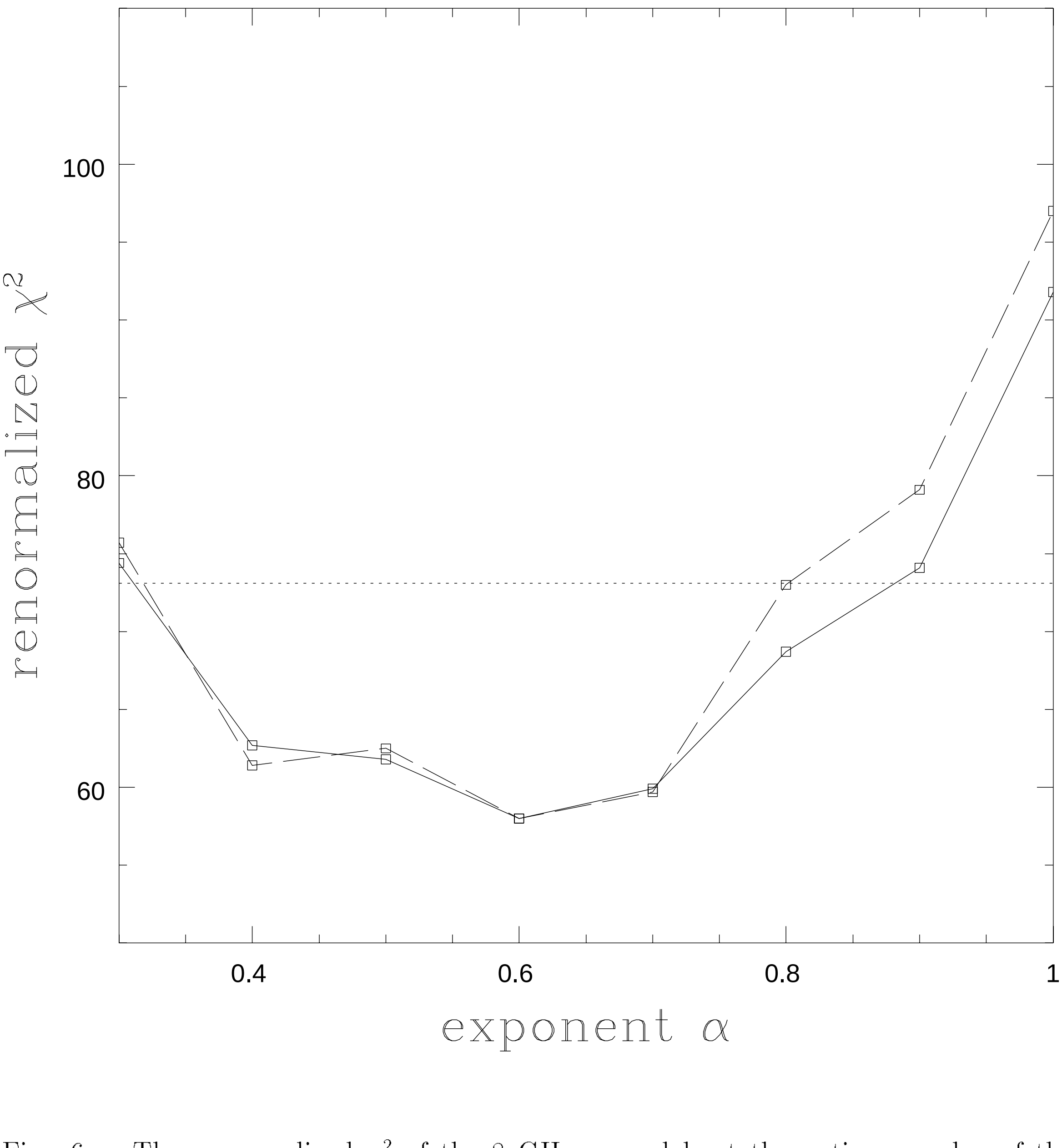

Fig. 6.— The renormalized $\chi^2$ of the 8 GHz $\alpha$ models at the optimum value of the core radius for each $\alpha$. The solid line shows the renormalized $\chi^2_{tot}$, and the dashed line shows the renormalized $\chi^2_{mult}$. The horizontal dotted line represents $\Delta\chi^2 = 15.1$ in the renormalized $\chi^2$ statistics.

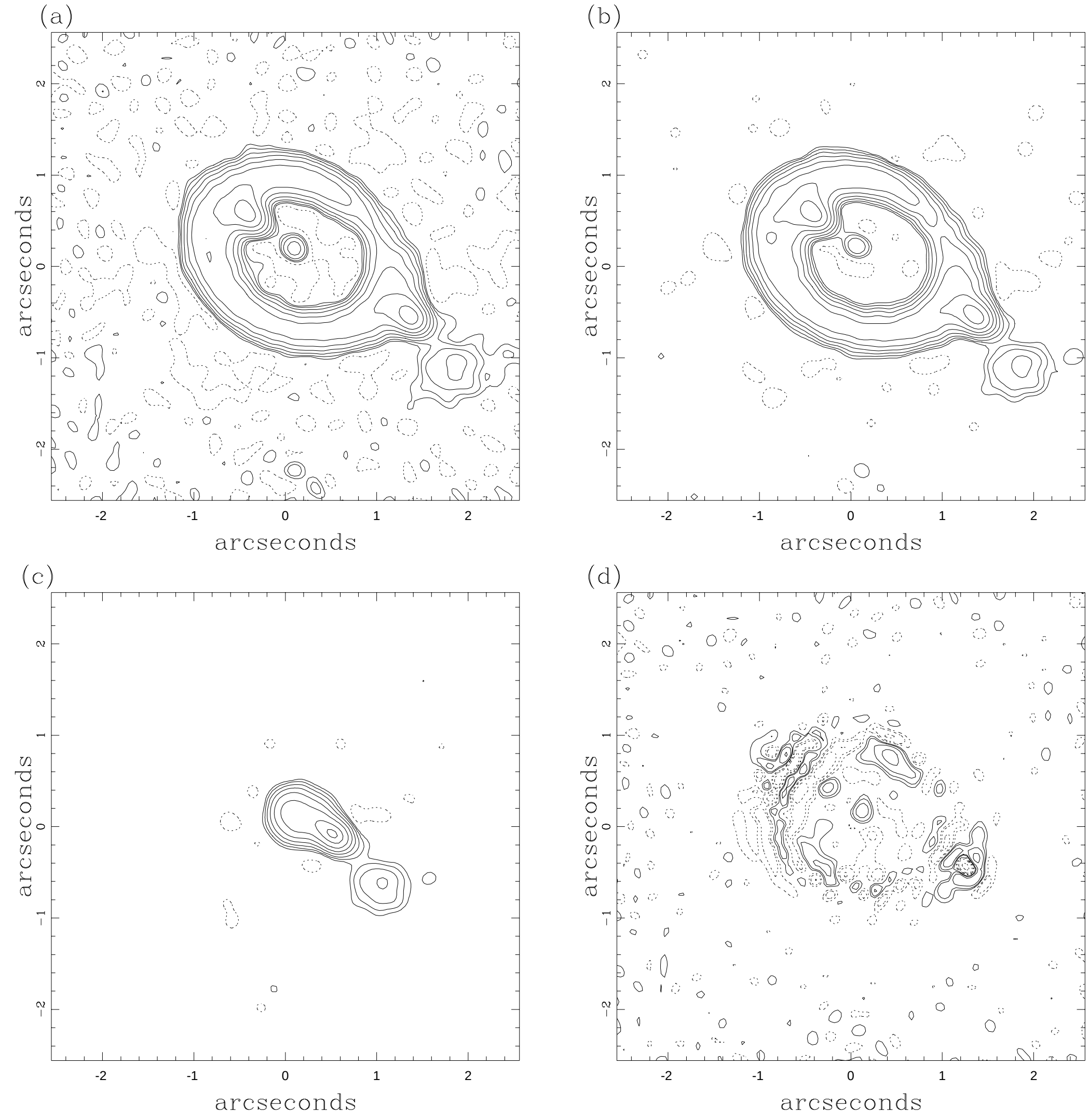

Fig. 7.— (a) The observed image, (b) the reconstructed image, (c) the inferred source, and (d) the residual map obtained from the best 8 GHz $\alpha$ model. The contour levels in the figures are: (a) and (b) $-1, 1, 2, 4, 8, 16, 32, 64, 128, 256 \times 35\mu$Jy, (c) $-1, 1, 2, 4, 8, 16, 32, 64, 95\% \times 155\mu$Jy, (d) $-8, -4, -2, -1, 1, 2, 4, 8 \times 35\mu$Jy. In a perfect reconstruction (a) and (b) would be identical.

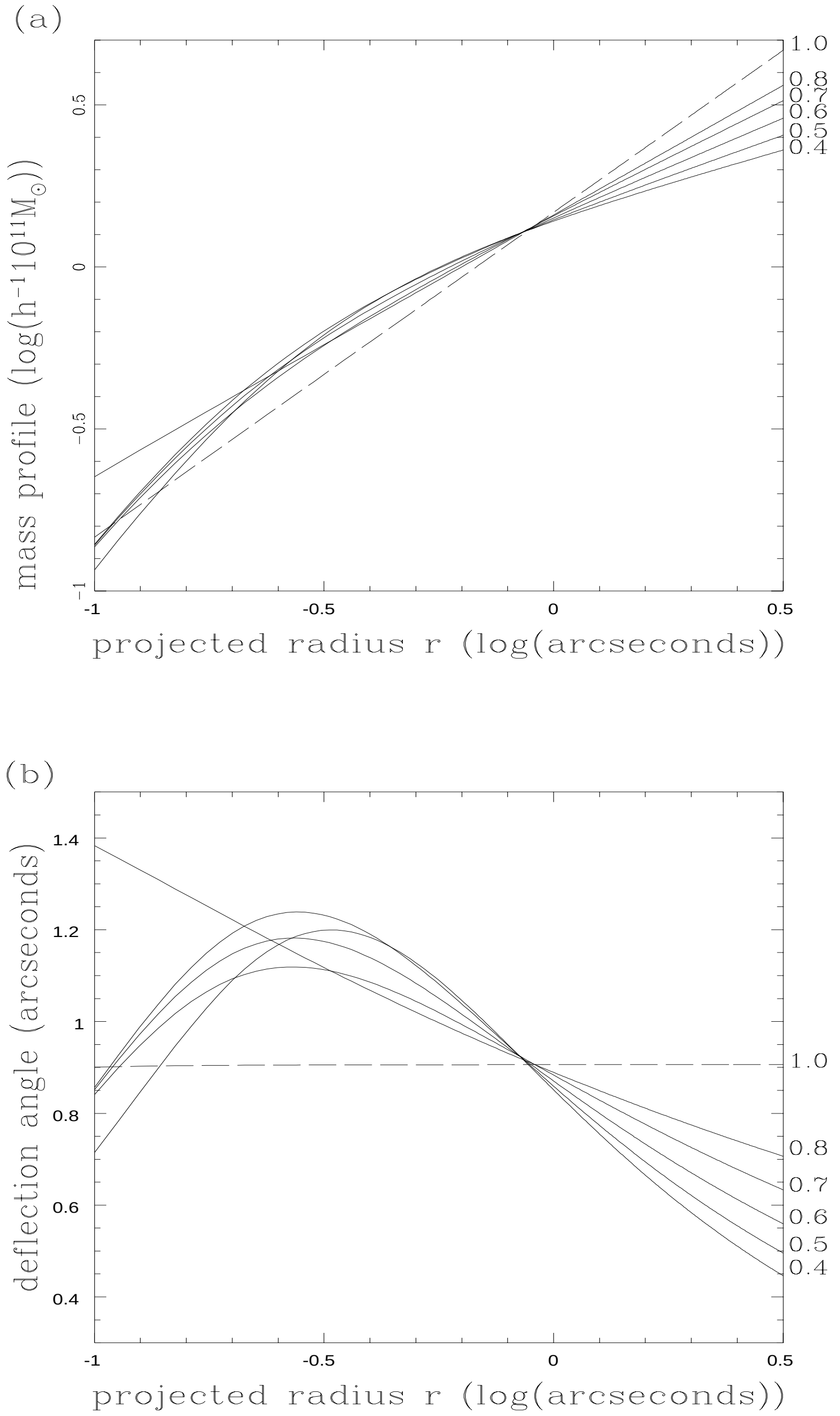

Fig. 8.— (a) The monopole mass distribution and (b) the monopole deflection of the acceptable 8 GHz $\alpha$ models as function of projected radius (solid lines). The mass is calculated assuming $z_s = 2.0$ and $z_l = 0.5$ in an Einstein de Sitter universe. For other values of redshifts, multiply the mass by $4.55 D_{os} D_{ol} / 2 r_h D_{ls}$, where $D_{ij}$ is the angular diameter distance between $i$ and $j$. The mass distribution and the monopole deflection of the best fit isothermal model are also shown for comparison (dashed lines).

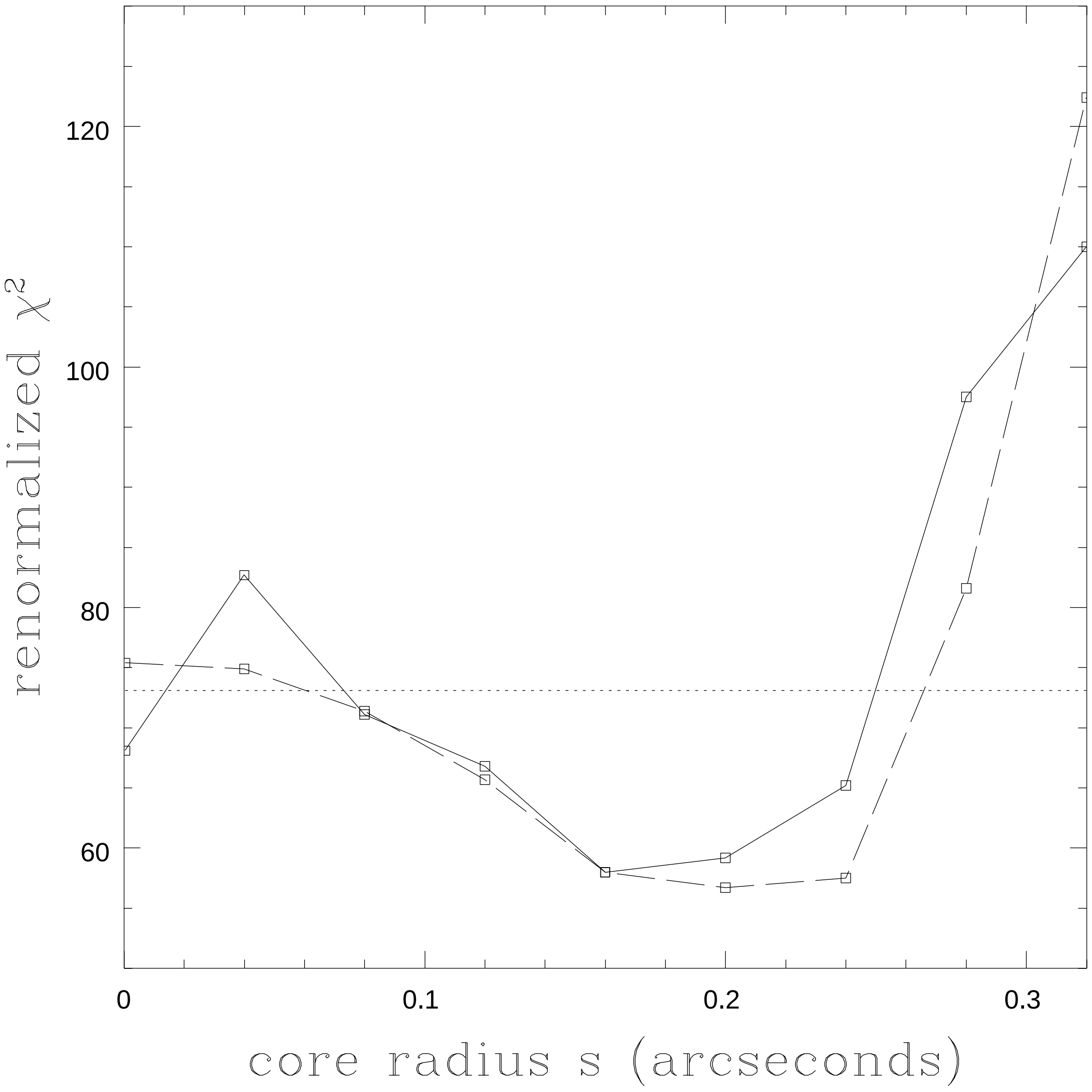

Fig. 9.— The results when D is treated as the radio image of the lensing galaxy. The dashed line shows the renormalized $\chi^2_{tot}$ as a function of the core radius at a fixed $\alpha$ ($\alpha = 0.6$) when D is subtracted before lens modeling. The results of the equivalent lens models when D is not subtracted are also shown (the solid line). The similarity of the statistics shows that the lens models are not affected by the nature of D. The dotted line shows $\Delta\chi^2_r = 15.1$ in the renormalized $\chi^2$ statistics.

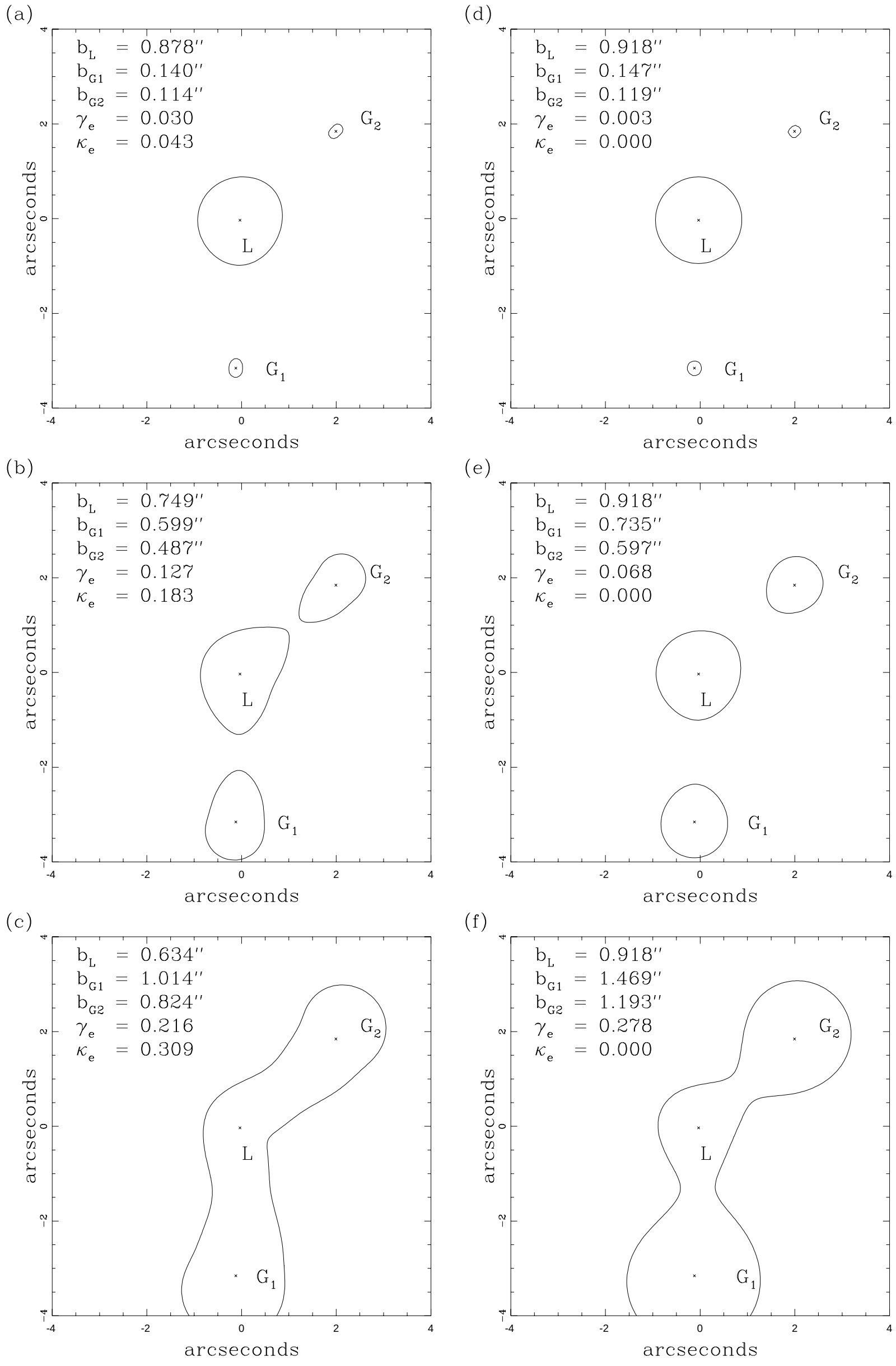

Fig. 10.— The critical lines produced by various combinations of $b_L$, $b_{G1}$, and $b_{G2}$ in an isothermal lens potential ((a) to (c)) and a point mass lens potential ((d) to (f)). The values of $b_L$, $b_{G1}$, and $b_{G2}$ are chosen so that, by using the Faber-Jackson relation, the ratios of the intrinsic luminosity between G1 and L (also G2 and L) are exactly as observed in (a) and (d), 25 times as observed (b) and (e), and 100 times as observed in (c) and (f). In all cases, the value of $b_L$ is chosen so that $b \sim 0\farcs918$ (the best fit value of the isothermal sphere model). The magnitudes of the shear ($\gamma_e$) and the convergence ($\kappa_e$) from G1 and G2 are also listed. If the masses of G1 and G2 are large enough, the magnitudes of the shear from G1 and G2 will exceed the value required to fit the ring ($\gamma = 0.13$).

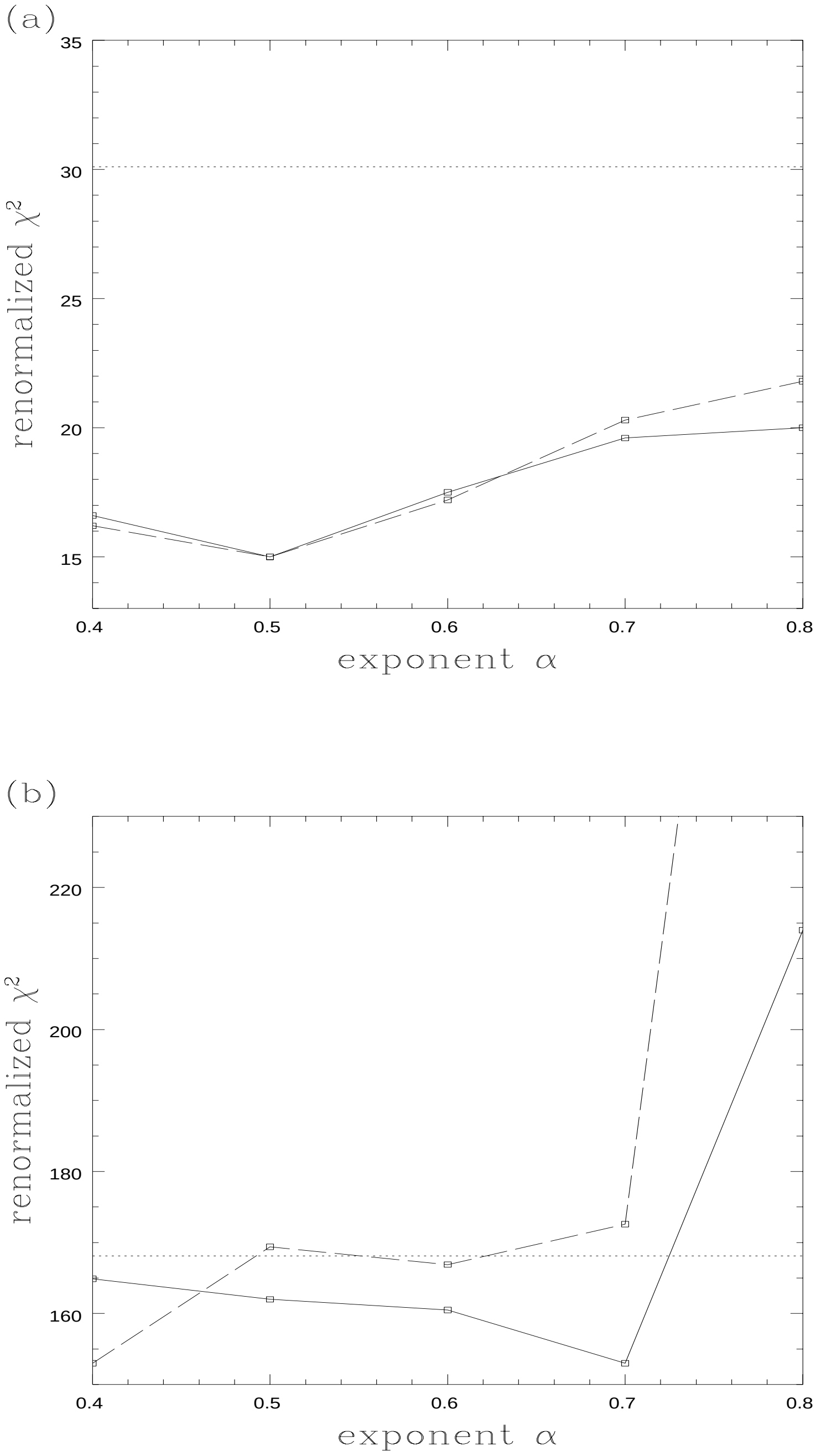

Fig. 11.— The renormalized $\chi^2$ as a function of $\alpha$ for the (a) 5 GHz and (b) 15 GHz $\alpha$ models. The solid line shows the renormalized $\chi^2_{tot}$ and the dashed line shows the renormalized $\chi^2_{mult}$. The dotted horizontal line shows $\Delta\chi^2 = 15.1$ in the renormalized $\chi^2$ statistics.

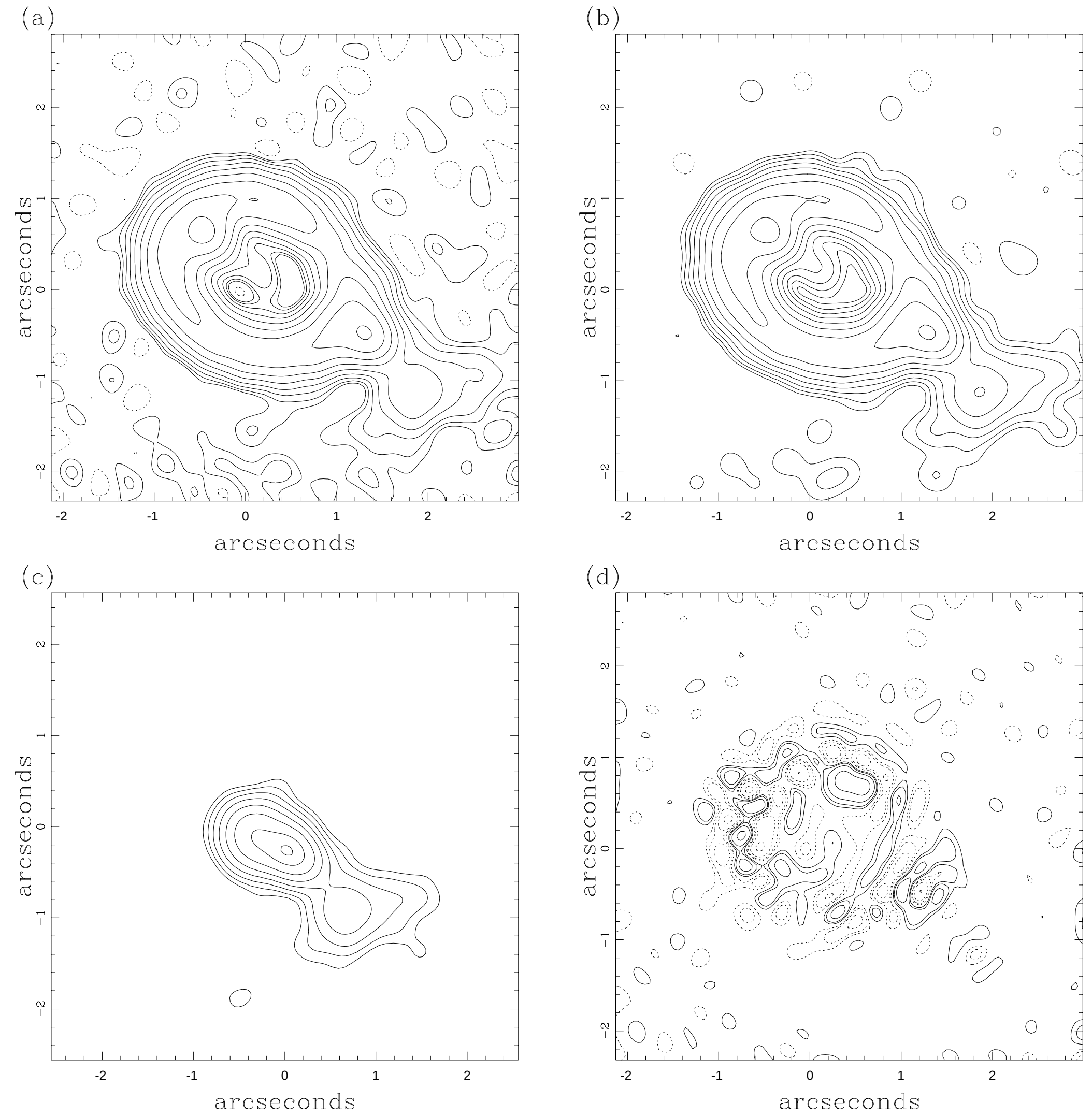

Fig. 12.— (a) The observed image, (b) the reconstructed image, (c) the inferred source, and (d) the residual map of the best 5 GHz $\alpha$ model. The contour levels are: (a) and (b) $-1, 1, 2, 4, 8, 16, 32, 64, 128, 256 \times 60\mu$Jy, (c) $-1, 1, 2, 4, 8, 16, 32, 64, 95\% \times 155\mu$Jy, (d) $-8, -4, -2, -1, 1, 2, 4, 8 \times 60\mu$Jy.

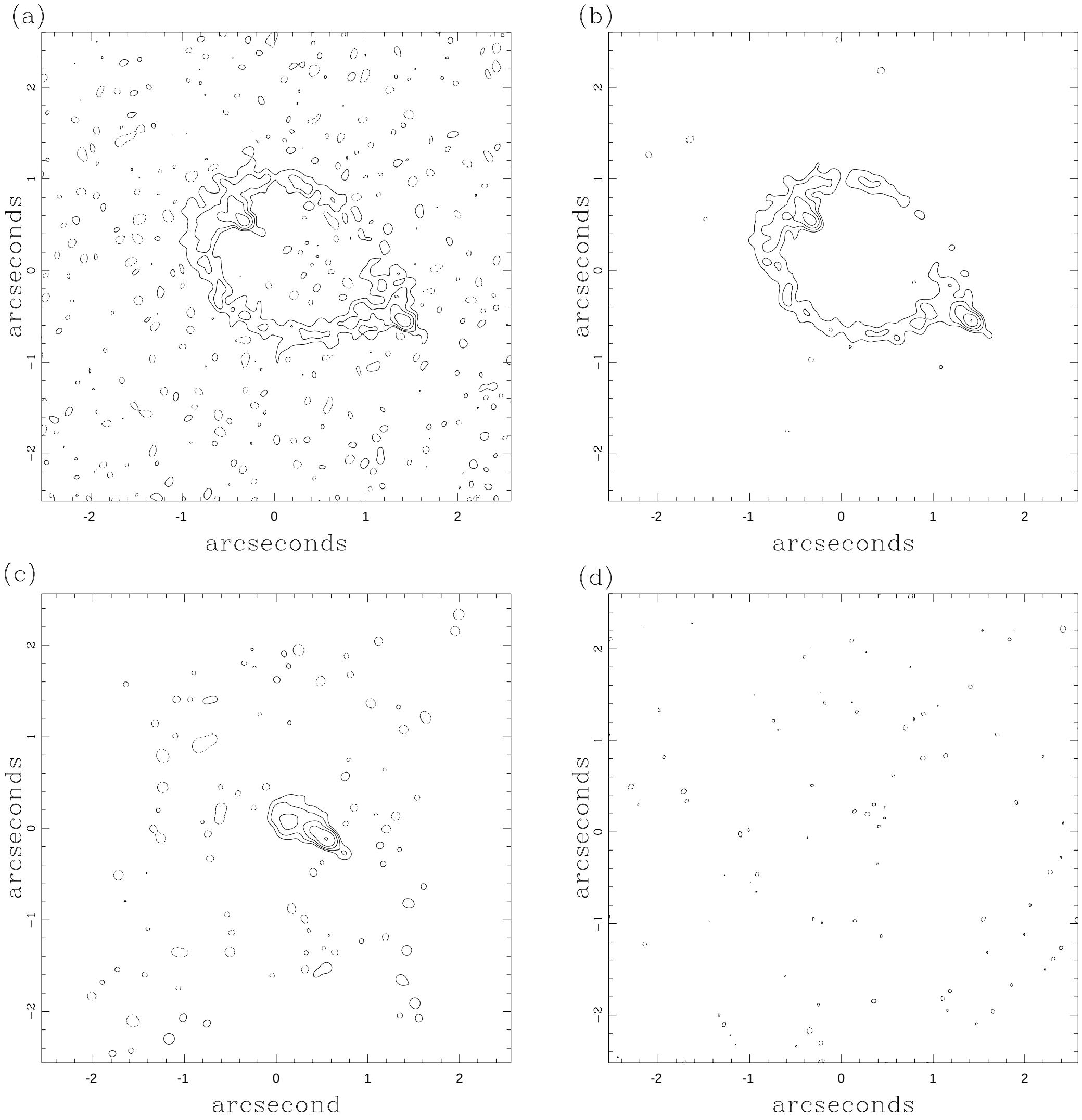

Fig. 13.— (a) The observed image, (b) the reconstructed image, (c) the inferred source, and (d) the residual map of the best 15 GHz $\alpha$ model. The contour levels are: (a) and (b) $-2, 2, 4, 8, 16, 32, 64, 128, 256 \times 130\mu$Jy, (c) $-4, 4, 8, 16, 32, 64, 95\% \times 62\mu$Jy, (d) $-4, -2, 2, 4 \times 130\mu$Jy.

Table 1. Summary of the Data

| frequency GHz | Synthesized Beam[a] | noise $\mu$Jy/pixel | peak mJy | pixel size ″/pixel | A[b] $(\pm0\farcs10)$[c] | B[b] $(\pm0\farcs10)$[c] | D[b] |
|---|---|---|---|---|---|---|---|
| 5 | $0\farcs33 \times 0\farcs32$ @ $-13°$ | 60 | 19.9 | 0.04 | (-0″.44, 0″.43) | (1″.20, -0″.76) | confused |
| 8 | $0\farcs19 \times 0\farcs19$ @ $-24°$ | 35 | 6.7 | 0.04 | (-0″.49, 0″.42) | (1″.21, -0″.75) | (0″.00, 0″.00) |
| 15 | $0\farcs12 \times 0\farcs12$ @ $-31°$ | 130 | 4.2 | 0.02 | (-0″.43, 0″.32) | (1″.31, -0″.76) | not detected |
| 22 | $0\farcs089 \times 0\farcs086$ @ $-50°$ | 200 | 3.8 | 0.02 | (-0″.43, 0″.32) | (1″.31, -0″.76) | not detected |

[a]FWHM of the major and minor axes of the synthesized beam, and its position angle measure north to east.

[b]The positions are given relative to $11^h31^m56.448^s + 04°55'49\farcs45$, the position of D at 8 GHz. The 5 and 8 GHz positions are determined using the Q maps, and the 15 and 22 GHz positions are determined using the I maps (see Chen & Hewitt 1993).

[c]The positional uncertainty given here is the the uncertainty in the absolute position. The uncertainty in relative positions measured on the same map is much smaller; we estimate that uncertainty to be 0″.01.

Table 2. Summary of the Data

| junk |
| --- |
| 0 |

Table 3. Summary of the Second Gain Experiment – Reoptimized Models

| gain | $b$[a] | $x_l$[b] | $y_l$[b] | $\gamma$[c] | $\theta_\gamma$[d] | $\sigma$[e] | $\Delta\sigma$[f] |
|---|---|---|---|---|---|---|---|
| | $''$ | $''$ | $''$ | | degrees | $\mu$Jy/pixel | $\mu$Jy |
| 0.20 | 0.923 | 0.007 | -0.004 | 0.125 | $-26.6$ | 40.7 | 368 |
| 0.15 | 0.927 | 0.009 | -0.014 | 0.124 | $-25.4$ | 41.7 | 356 |
| 0.10 | 0.919 | 0.009 | -0.006 | 0.125 | $-26.2$ | 37.6 | 343 |
| 0.08 | 0.921 | 0.008 | -0.006 | 0.123 | $-26.3$ | 40.4 | 393 |
| 0.06 | 0.921 | 0.010 | -0.008 | 0.124 | $-26.1$ | 38.7 | 363 |
| 0.05 | 0.922 | 0.008 | -0.007 | 0.125 | $-25.9$ | 39.6 | 375 |
| 0.02 | 0.919 | 0.002 | -0.005 | 0.127 | $-25.9$ | 40.0 | 342 |
| mean | $0.922 \pm 0.003$ | $0.008 \pm 0.003$ | $-0.007 \pm 0.003$ | $0.125 \pm 0.001$ | $-26.1 \pm 0.4$ | $39.8 \pm 1.3$ | $363 \pm 18$ |

[a]The critical radius.

[b]The position of the lens relative to component D.

[c]The dimensionless shear.

[d]The position angle of the shear (measured from west to north).

[e]The rms flux in the residual map.

[f]The peak flux in the residual map.

Table 4. The 8 GHz de Vaucouleurs Models

| $R_e$ | $\Delta\sigma$ | $\chi^2_{tot}$ [a] | $\chi^2_{mult}$ [a] | $b$ | $x_l$ [b] | $y_l$ [b] | $\gamma$ | $\theta_\gamma$ |
|---|---|---|---|---|---|---|---|---|
| $''$ | $\mu$Jy | | | $''$ | $''$ | $''$ | | degrees |
| 0.56 | 436 | 1244 | 1100 | $1.094 \pm 0.012$ | $-0.017 \pm 0.006$ | $0.003 \pm 0.008$ | $0.164 \pm 0.006$ | $-26.0 \pm 0.4$ |
| 0.60 | 461 | 1173 | 1002 | $1.053 \pm 0.005$ | $-0.016 \pm 0.007$ | $0.002 \pm 0.011$ | $0.158 \pm 0.006$ | $-26.1 \pm 0.7$ |
| 0.64 | 446 | 1066 | 899 | $1.032 \pm 0.015$ | $-0.011 \pm 0.012$ | $0.010 \pm 0.013$ | $0.156 \pm 0.008$ | $-26.5 \pm 0.9$ |
| 0.72 | 393 | 880 | 743 | $0.995 \pm 0.013$ | $-0.003 \pm 0.010$ | $-0.009 \pm 0.010$ | $0.143 \pm 0.010$ | $-25.5 \pm 1.0$ |
| 0.80 | 366 | 889 | 741 | $0.950 \pm 0.011$ | $-0.001 \pm 0.011$ | $-0.005 \pm 0.017$ | $0.141 \pm 0.009$ | $-25.9 \pm 1.2$ |
| 0.88 | 390 | 987 | 837 | $0.922 \pm 0.010$ | $0.011 \pm 0.010$ | $-0.014 \pm 0.010$ | $0.129 \pm 0.006$ | $-25.9 \pm 1.6$ |
| 0.96 | 404 | 1056 | 896 | $0.893 \pm 0.008$ | $0.019 \pm 0.009$ | $-0.017 \pm 0.013$ | $0.128 \pm 0.011$ | $-26.1 \pm 2.0$ |
| 1.00 | 467 | 1285 | 1106 | $0.879 \pm 0.018$ | $0.023 \pm 0.011$ | $-0.017 \pm 0.025$ | $0.120 \pm 0.007$ | $-25.5 \pm 1.3$ |
| 1.04 | 519 | 1664 | 1475 | $0.885 \pm 0.012$ | $0.025 \pm 0.011$ | $-0.020 \pm 0.011$ | $0.114 \pm 0.007$ | $-24.9 \pm 1.7$ |
| $0.83 \pm 0.13$ | 342 | 825 | 703 | $0.941 \pm 0.034$ | $0.005 \pm 0.014$ | $-0.010 \pm 0.011$ | $0.135 \pm 0.015$ | $-25.9 \pm 0.6$ |

[a] $N_{dof} = 57$.

[b] The position of the lens is given relative to the position of component D at 8 GHz.

Table 5. The 8 GH $\alpha$ Models

| $\alpha$ | $\Delta\sigma$ | $\chi^2_{tot}$ [a] | $\chi^2_{mult}$ [a] | $b$ | $x_l$ [b] | $y_l$ [b] | $\gamma$ | $\theta_\gamma$ | $s$ |
|---|---|---|---|---|---|---|---|---|---|
| | $\mu$Jy | | | $''$ | $''$ | $''$ | | degrees | $''$ |
| 0.3 | 398 | 955 | 797 | $0.951 \pm 0.039$ | $0.015 \pm 0.011$ | $-0.007 \pm 0.018$ | $0.133 \pm 0.005$ | $-26.5 \pm 1.9$ | $0.290 \pm 0.090$ |
| 0.4 | 386 | 804 | 646 | $0.933 \pm 0.022$ | $0.001 \pm 0.022$ | $-0.007 \pm 0.018$ | $0.137 \pm 0.009$ | $-25.7 \pm 1.7$ | $0.254 \pm 0.065$ |
| 0.5 | 344 | 792 | 658 | $0.922 \pm 0.025$ | $0.004 \pm 0.010$ | $-0.011 \pm 0.018$ | $0.132 \pm 0.006$ | $-25.8 \pm 1.8$ | $0.196 \pm 0.064$ |
| 0.6 | 343 | 744 | 611 | $0.919 \pm 0.008$ | $0.009 \pm 0.010$ | $-0.006 \pm 0.007$ | $0.125 \pm 0.003$ | $-26.2 \pm 0.7$ | $0.172 \pm 0.068$ |
| 0.7 | 346 | 768 | 628 | $0.919 \pm 0.015$ | $-0.001 \pm 0.012$ | $0.002 \pm 0.019$ | $0.123 \pm 0.008$ | $-26.5 \pm 1.5$ | $0.148 \pm 0.065$ |
| 0.8 | 421 | 880 | 768 | $0.907 \pm 0.016$ | $0.001 \pm 0.007$ | $0.008 \pm 0.025$ | $0.113 \pm 0.006$ | $-27.2 \pm 2.0$ | $0.018 \pm 0.101$ |
| 0.9 | 366 | 951 | 832 | $0.912 \pm 0.018$ | $-0.015 \pm 0.018$ | $-0.001 \pm 0.023$ | $0.111 \pm 0.007$ | $-25.6 \pm 1.6$ | $0.024 \pm 0.096$ |
| 1.0 | 411 | 1178 | 1022 | $0.918 \pm 0.011$ | $-0.020 \pm 0.023$ | $0.017 \pm 0.022$ | $0.107 \pm 0.010$ | $-27.2 \pm 2.5$ | $0.026 \pm 0.074$ |
| $0.63 \pm 0.23$ | 357 | 740 | 598 | $0.924 \pm 0.017$ | $-0.002 \pm 0.007$ | $-0.002 \pm 0.010$ | $0.124 \pm 0.012$ | $-25.9 \pm 1.3$ | $0.187 \pm 0.067$ |

[a] $N_{dof} = 58$.

[b] The position of the lens is given relative to the position of component D at 8 GHz.

Table 6. The 8 GH $\alpha$ Models

| junk |
|---|
| 0 |

Table 7. The 5 GHz $\alpha$ Models

| $\alpha$ | $\Delta\sigma$ | $\chi^2_{tot}$ [a] | $\chi^2_{mult}$ [a] | $b$ | $x_l$ [b] | $y_l$ [b] | $\gamma$ | $\theta_\gamma$ | $s$ |
|---|---|---|---|---|---|---|---|---|---|
| | $\mu$Jy | | | $''$ | $''$ | $''$ | | degrees | $''$ |
| 0.4 | 639 | 449 | 392 | $0.936 \pm 0.003$ | $0.020 \pm 0.016$ | $0.007 \pm 0.011$ | $0.145 \pm 0.011$ | $-25.9 \pm 1.1$ | $0.258 \pm 0.004$ |
| 0.5 | 605 | 407 | 363 | $0.921 \pm 0.005$ | $0.022 \pm 0.012$ | $0.004 \pm 0.014$ | $0.133 \pm 0.015$ | $-25.8 \pm 1.7$ | $0.196 \pm 0.004$ |
| 0.6 | 682 | 473 | 418 | $0.921 \pm 0.005$ | $0.026 \pm 0.020$ | $0.005 \pm 0.012$ | $0.130 \pm 0.011$ | $-26.1 \pm 1.7$ | $0.162 \pm 0.018$ |
| 0.7 | 707 | 531 | 481 | $0.916 \pm 0.009$ | $0.022 \pm 0.017$ | $0.009 \pm 0.022$ | $0.125 \pm 0.012$ | $-26.4 \pm 2.6$ | $0.120 \pm 0.029$ |
| 0.8 | 739 | 541 | 526 | $0.909 \pm 0.006$ | $0.019 \pm 0.022$ | $0.015 \pm 0.012$ | $0.115 \pm 0.009$ | $-26.3 \pm 1.8$ | $0.018 \pm 0.004$ |
| $0.5 \pm 0.3$ | 605 | 407 | 363 | $0.921 \pm 0.016$ | $0.022 \pm 0.012$ | $0.004 \pm 0.015$ | $0.133 \pm 0.018$ | $-25.8 \pm 0.6$ | $0.196 \pm 0.058$ |

[a] $N_{dof} = 15$.

[b] The position of the lens is given relative to the position of component D at 8 GHz.

Table 8. The 15 GH $\alpha$ Models

| $\alpha$ | $\Delta\sigma$ | $\chi^2_{tot}$ [a] | $\chi^2_{mult}$ [a] | $b$ | $x_l$ [b] | $y_l$ [b] | $\gamma$ | $\theta_\gamma$ | $s$ |
|---|---|---|---|---|---|---|---|---|---|
| | $\mu$Jy | | | $''$ | $''$ | $''$ | | degrees | $''$ |
| 0.4 | 559 | 1128 | 379 | $0.938 \pm 0.004$ | $0.024 \pm 0.004$ | $0.004 \pm 0.003$ | $0.133 \pm 0.006$ | $-26.6 \pm 0.4$ | $0.254 \pm 0.003$ |
| 0.5 | 597 | 1107 | 420 | $0.922 \pm 0.008$ | $0.021 \pm 0.006$ | $0.002 \pm 0.005$ | $0.132 \pm 0.011$ | $-25.8 \pm 0.8$ | $0.196 \pm 0.005$ |
| 0.6 | 421 | 1059 | 403 | $0.925 \pm 0.005$ | $0.049 \pm 0.008$ | $-0.015 \pm 0.007$ | $0.115 \pm 0.011$ | $-26.5 \pm 0.6$ | $0.172 \pm 0.008$ |
| 0.7 | 467 | 1047 | 428 | $0.914 \pm 0.008$ | $0.022 \pm 0.004$ | $0.003 \pm 0.003$ | $0.125 \pm 0.002$ | $-26.5 \pm 1.1$ | $0.148 \pm 0.005$ |
| 0.8 | 540 | 1465 | 902 | $0.907 \pm 0.010$ | $0.021 \pm 0.010$ | $0.002 \pm 0.010$ | $0.113 \pm 0.009$ | $-27.2 \pm 2.2$ | $0.018 \pm 0.010$ |
| $0.7 \pm 0.2$ | 467 | 1047 | 427 | $0.914 \pm 0.010$ | $0.022 \pm 0.027$ | $0.003 \pm 0.018$ | $0.125 \pm 0.010$ | $-26.5 \pm 0.6$ | $0.145 \pm 0.048$ |

[a] $N_{dof} = 153$.

[b] The position of the lens is given relative to the position of component D at 8 GHz.